%% file: rollercoasters.tex
\documentclass[a4paper,UKenglish]{lipics-v2021}
\usepackage{mathtools,booktabs,tikz,needspace,placeins,xifthen}
\usetikzlibrary{patterns,decorations.pathreplacing,arrows.meta,shapes.geometric,shapes.multipart,positioning}
\hideLIPIcs

\nolinenumbers

\title{Exact guarantees and faster algorithms for \texorpdfstring{$k$}{k}-rollercoasters}
\titlerunning{Exact guarantees and faster algorithms for k-rollercoasters}
\author{Sungmin Kim}{Department of Informatics, University of G{\"o}ttingen, Germany}{sungmin.kim@cs.uni-goettingen.de}{0000-0003-3153-0314}{}
\author{Florin Manea}{Department of Informatics, University of G{\"o}ttingen, Germany}{florin.manea@cs.uni-goettingen.de}{0000-0001-6094-3324}{}
\authorrunning{Kim and Manea}
\Copyright{Kim and Manea}

\newcommand{\LIS}{\operatorname{LIS}}
\newcommand{\LDS}{\operatorname{LDS}}
\newcommand{\rc}{\operatorname{rc}}
\newcommand{\ceil}[1]{\left\lceil#1\right\rceil}
\newcommand{\floor}[1]{\left\lfloor#1\right\rfloor}
\newcommand{\inc}{\prec_{+}}
\newcommand{\dec}{\prec_{-}}
\newcommand{\Zend}{Z_{\mathrm{end}}}
\newcommand{\Zstart}{Z_{\mathrm{start}}}
\newcommand{\core}[1]{{#1}^{\circ}}   % core of a permutation S: S minus Zend and Zstart
\newcommand{\estrand}{\mathsf{es}}   % ending strand of a point in a three-layer representation
\newcommand{\sstrand}{\mathsf{ss}}   % starting strand
\NewDocumentCommand{\layerrep}{g g}{three-layer\IfValueTF{#1}{ $(#1,#2)$-}{ }representation}   % three-layer (c,w)-representation; \layerrep alone prints "three-layer representation"
\newcommand{\strandc}[1]{\textup{(S#1)}}   % tag (S1)..(S4) of a strand constraint in Definition def:layers, always upright
\NewDocumentCommand{\ie}{m g}{\mathsf{ie}_{#1}\IfValueT{#2}{(#2)}}   % increasing chain ending at the point
\NewDocumentCommand{\de}{m g}{\mathsf{de}_{#1}\IfValueT{#2}{(#2)}}   % decreasing chain ending at the point
\NewDocumentCommand{\is}{m g}{\mathsf{is}_{#1}\IfValueT{#2}{(#2)}}   % increasing chain starting at the point
\NewDocumentCommand{\ds}{m g}{\mathsf{ds}_{#1}\IfValueT{#2}{(#2)}}   % decreasing chain starting at the point
\newcommand{\extremalf}[2]{f_{#1}(#2)}                             % f_k(n), guaranteed rollercoaster length
\newcommand{\extremalN}[2]{N_{#1}(#2)}                             % N_k(L), largest input with rc_k <= L
\newcommand{\rclen}[2]{\rc_{#1}(#2)}                               % rc_k(S), longest k-rollercoaster in S
\newcommand{\rcend}[1]{e(#1)}                                      % e(p), longest rollercoaster ending at p
\newcommand{\rcstart}[1]{s(#1)}                                    % s(p), longest rollercoaster starting at p
\newcommand{\incclass}[2]{\mathcal I_{#1}(#2)}                     % points of S with increasing end rank i
\newcommand{\decclass}[2]{\mathcal D_{#1}(#2)}                     % points of S with decreasing end rank j
\newcommand{\gridperm}[2]{R(#1,#2)}                                % R(h,v), h decreasing blocks of v points
\newcommand{\sevenperm}[2]{W_{#1,#2}}                              % W_{q,t}, the seven-block permutation
\newcommand{\Ustar}[1]{U^{*}_{#1}}                                 % a largest of the sets U_1,...,U_i; |\Ustar{i}| is the prefix maximum of their sizes
\newcommand{\Dstar}[1]{D^{*}_{#1}}                                 % a largest of the sets D_1,...,D_j
\newcommand{\startlen}[1]{g_0(#1)}                                 % starting length g_0(v) prescribed at a start position of a scan
\newcommand{\scanlen}[1]{g(#1)}                                    % assigned length g(p) computed by a scan
\newcommand{\class}[1]{H_{#1}}                                     % classes H_h of Algorithm 1
\newcommand{\stored}[1]{\rho(#1)}                                  % stored rollercoaster rho(p)
\newcommand{\segidx}[1]{\lambda(#1)}                               % index lambda(p) of the segment containing p
\newcommand{\dptab}[2][]{F_{#2}\ifthenelse{\isempty{#1}}{}{(#1)}}  % table F_sigma; \dptab[j,s]{\sigma} is the entry F_sigma(j,s)
\newcommand{\dppred}[3]{\omega_{#1}(#2,#3)}                        % sigma-predecessor weight omega_sigma(p,s)
\newcommand{\pot}{\Phi}                                            % potential of the finger sets

\begin{document}
\maketitle
\begin{abstract}
For $k\ge3$, a $k$-rollercoaster is a sequence of distinct values whose
maximal increasing and decreasing runs all have length at least $k$.
We discover the maximum guaranteed length $\extremalf{k}{n}$ of $k$-rollercoasters
appearing as subsequences of $n$-point permutations.
In particular, letting $q=k-1$, we prove that
$\extremalf{k}{n}= \ceil{n/q}$ for $n\ge2q^2$, 
$\extremalf{k}{n}= 2q-\lfloor \sqrt{2q^2-n}\rfloor$ for $q^2<n<2q^2$
and $\extremalf{k}{n}= 0$ otherwise. We prove that the guarantees are tight
by constructing an $n$-point permutation
whose longest $k$-rollercoaster has exactly $\extremalf{k}{n}$ points.
Our proof extends
Seidenberg's argument~[Journal of the London Mathematical Society, 1959] for proving the Erd\H{o}s--Szekeres
theorem
by representing points in a subsequence with respect to
the lengths of longest monotonic (increasing and decreasing) subsequences
that reach each point.

The combinatorial proof yields a deterministic comparison-based
algorithm that finds a rollercoaster of length at least $\extremalf{k}{n}$
in $O(n\log \extremalf{k}{n})$ time and $O(n)$ space for $n>q^2$. Independently,
we give an $O(nk\log k)$-time, $O(nk)$-space algorithm for a longest
$k$-rollercoaster, improving the previous $O(nk^2)$ bound. We use the
dynamic program of Gawrychowski et al.~[STACS 2019, Algorithmica 2022],
and process
its weighted predecessor queries through monotone subsequences with
finger-search trees. Both algorithms return such subsequences.
\end{abstract}

\paragraph*{Disclosure of AI assistance.}
ChatGPT 6 Astra and Claude Fable 5.1 assisted in developing the extremal functions, sharp
constructions, combinatorial argument, and both algorithms, and in
checking intermediate proofs and examples. They also assisted with
expository revision and figures. Their contributions included
mathematical development. Complete proofs are included below; the two named
authors retain responsibility for the submitted work.

\addtocounter{page}{-1}
\clearpage

\input{first-12-pages.tex}

\FloatBarrier
\clearpage
\bibliography{rollercoasters}
\clearpage
\appendix

\input{appendix.tex}

% END OPTIONAL SUPPLEMENT

\end{document}

%% file: first-12-pages.tex
\section{Introduction}\label{sec:intro}
A \emph{subsequence} of $S=(x_1,\ldots,x_n)$ is a sequence
$(x_{i_1},\ldots,x_{i_\ell})$ with $1\le i_1<\cdots<i_\ell\le n$
that is not necessarily consecutive in $S$.
We consider sequences with distinct values, so only their relative order
matters, as in permutation pattern matching~\cite{BulteauFJV22}.
A permutation occurs as a \emph{pattern} in $S$ if some subsequence of $S$
has the same relative order as that permutation.

The study of patterns occurring in permutations as subsequences has a
long tradition; see, for example, Kitaev's monograph~\cite{Kitaev11}.
Especially, the longest increasing subsequence (LIS) and the longest
decreasing subsequence (LDS) are important examples of permutations
as subsequences. The well-established Erd\H{o}s--Szekeres
theorem~\cite{ErdosS35} guarantees a monotone subsequence of
length at least $\ceil{\sqrt n}$ in every permutation of length $n$,
and this bound is tight. Furthermore, Dilworth's theorem~\cite{Dilworth50}
says that the minimum number of increasing subsequences partitioning
a permutation equals its longest decreasing subsequence length,
and symmetrically for decreasing subsequences.

We study subsequences that may change direction, but only after progressing
monotonically for at least a given number of steps in one direction.
Such a maximal contiguous increasing or decreasing part
of a sequence is called a \emph{run}, where adjacent runs share the
turning point. Then, for $k\ge3$,
a \emph{$k$-rollercoaster} is a sequence of length at least $k$
in which every run has at least $k$ points. Note that this definition
allows a single
monotone run. For example, $(1,3,5,7,6,4,2)$ is a $4$-rollercoaster, with
two runs sharing $7$. Throughout this paper, we refer to a $k$-rollercoaster
simply by rollercoaster when the context is clear.
We analyse rollercoasters occurring as subsequences of permutations,
thus seeking long patterns of a given form.

The motivation for studying rollercoasters for $k=3$ arose in drawing
trees on given
point sets with L-shaped edges~\cite{BiedlCDJL18}. Here, Biedl et
al.~\cite{BiedlBCLMNS18,BiedlBCLMNS19} proved that every permutation of $n\ge8$
points contains a $3$-rollercoaster of length $\ceil{n/2}$, and that
this is the best length guarantee. Compared to the Erd\H{o}s--Szekeres
theorem, the length guarantee increases from order $\sqrt n$ to order $n$
when we allow changes in direction.

Our main goal is to generalise this exact guarantee for $k=3$ to
arbitrary $k\ge3$. Let us write $\rclen{k}{S}$ for the length of the
longest rollercoaster in a given permutation $S$,
where $\rclen{k}{S}=0$ if none exists. Let
$q=k-1$, and let
$\extremalf{k}{n}$ be the minimum value of $\rclen{k}{S}$
over all permutations~$S$ of $[n]=\{1,\ldots,n\}$.
Thus the extremal~$\extremalf{k}{n}$ is a worst-case
guarantee for any sequence of length~$n$,
while $\rclen{k}{S}$ is the optimum for one given sequence.
For $k\ge4$, the bounds of Biedl et al.~\cite{BiedlBCLMNS18} were
$\frac{n}{2q}-\frac{3(q+1)}{2}\le \extremalf{k}{n}\le\ceil{n/q}$
for the case when $n>q^2$.
We identify the value of $\extremalf{k}{n}$ for every $n$,
and we observe that
the previous upper-bound $\ceil{n/q}$ was tight
for every $n\ge2q^2$.

\begin{theorem}\label{thm:main}
For every $k\ge3$, with $q=k-1$, and every integer $n\ge0$,
\[
 \extremalf{k}{n}=
 \begin{cases}
 0,&0\le n\le q^2,\\
 2q-\floor{\sqrt{2q^2-n}},&q^2<n<2q^2,\\
 \ceil{n/q},&n\ge2q^2.
 \end{cases}
\]
Moreover, for every $n\ge0$, there is an explicitly constructed
$n$-point permutation $S$ with $\rclen{k}{S}=\extremalf{k}{n}$.
\end{theorem}

For $k=3$, the theorem agrees with the known answer; for $k\ge4$, it
closes the gap in the bound for $\extremalf{k}{n}$.
For example, $f_4(n)$ is zero up to $n=9$,
four for $10\le n\le14$, five for $15\le n\le17$, and $\ceil{n/3}$
from $n=18$ onwards. The construction at $n=17$ is drawn in
Figure~\ref{fig:block-template}.

\subparagraph*{Proof strategy.}\label{sec:strategy}
We extend Seidenberg's rank-pair proof~\cite{Seidenberg59} of the
Erd\H{o}s--Szekeres theorem. A point's increasing and decreasing
\emph{ending ranks} are the longest lengths of
increasing and decreasing subsequences ending there, respectively.
One rank increases between any earlier and later point,
so all pairs are distinct. Starting ranks are defined analogously.

Our approach is to identify the dual of the extremal function~$\extremalf{k}{n}$.
Specifically, we bound the length of the permutation
whose rollercoaster subsequences have length at most $L$.
For $L\le2q$, the number of different ending and starting rank-pairs
gives the exact count: $q^2$ for $0\le L\le q$, and
$2q^2-(2q-L)^2$ for $q\le L\le2q$. For larger $L$, we repeatedly
delete at most $q^2$ points and reduce $L$ by $q$, until
$L=2q+r$ with $0\le r<q$. Then, for the remaining sequence~$S$,
at most $2q^2$ remaining points cannot be
an end and a start of a rollercoaster at the same time,
whereas the points that can be both form a
subsequence $\core{S}$ with no monotone subsequence longer than $q$.

If we directly apply the Erd\H{o}s--Szekeres theorem,
we have $|\core{S}|\le q^2$. However we prove
a tighter bound $|\core{S}|\le q\min(q,r)$
by representing $\core{S}$ in terms of the surrounding
permutation~$S$. At every point of $\core{S}$ in the sequence~$S$,
a monotone subsequence of $S$ with at least $q+1$
points ends and another starts.
Take a longest increasing or decreasing subsequence
that starts or ends at a point.
Joining starting and ending subsequences at the point gives a rollercoaster, so
their lengths sum to at most $2q+r+1$. Write the lengths at the point $p$ as $q+\estrand(p)$ and $q+\sstrand(p)$.
The strictly positive strands then satisfy $\estrand(p)+\sstrand(p)\le r+1$; in particular,
$\core{S}$ is empty when $r=0$.

We count the number of points in $\core{S}$
by reformulating $\core{S}$ into
a sequence~$P$ with three layers: an upper, a middle,
and a lower layer, where no union of two adjacent layers
contains an increasing subsequence with $r+1$ points.
The reformulation also requires
that the longest increasing and decreasing subsequences
have at most $q$ points.
We prove by induction on $q$
that any sequence~$P$ that can be represented in such a way
has at most $q\min(q,r)$ points.
Note that if $q\le r$,
we have $|P|\le q^2=q\min(q,r)$.

Consider the case when
$P$ has an increasing $q$-point subsequence.
We combine the subsequence
with a longest increasing subsequence in the middle layer
of length $h$. Cutting and rejoining gives two increasing
subsequences of total length $q+h$, each avoiding one outer layer,
and this gives $q+h\le2r$.
Using this bound,
we then count the three layers together
by bounding the number of upper-left points
and the lower-right points of $P$
each by $\frac{r(r+1)}{2}$,
which gives $|P|\le r(r+1)\le qr$.

For the remaining case when $P$
does not have an increasing $q$-point subsequence,
we partition $P$ into $q$ sets of increasing subsequences,
and delete a smallest class.
This removes at most $|P|/q$ points,
and the resulting sequence cannot have a decreasing subsequence
of length $q$.
Thus, this gives us another sequence
that satisfies the three-layer formation for $q-1$.
The induction hypothesis gives
$(q-1)|P|/q\le(q-1)r$, and hence $|P|\le qr$.

\subparagraph*{Our algorithmic results.}
For $k=3$, Biedl et al.'s proof of the sharp guarantee is constructive:
it finds a $3$-rollercoaster of length $\ceil{n/2}$ in linear time
for $n\ge8$~\cite{BiedlBCLMNS18,BiedlBCLMNS19}. For general $k$, their method
finds a rollercoaster meeting their weaker guarantee
of $\frac{n}{2q}-\frac{3(q+1)}{2}$ in
$O(n\log k)$ time~\cite{GawrychowskiMS22}.
Our combinatorial proof also yields an algorithm, which finds a
rollercoaster subsequence of length at least the optimal guarantee $\extremalf{k}{n}$ of
Theorem~\ref{thm:main}. This byproduct of the proof takes $O(n\log \extremalf{k}{n})$ time and $O(n)$ space for $n>q^2$
(Theorem~\ref{thm:algorithm}). 
%It may be slower than the earlier
%algorithm, although the bounds have the same order when
%$n=q^{O(1)}$.
The returned rollercoaster is not necessarily the longest in the
given permutation.

We also revisit the longest-rollercoaster
algorithm of Gawrychowski et al.~\cite{GawrychowskiMS19,GawrychowskiMS22}.
We improve their $O(nk^2)$ running time to
$O(nk\log k)$, using $O(nk)$ space (Theorem~\ref{opt:thm:longest}).
We retain their decomposition and dynamic program, and replace the
data structure for its weighted predecessor queries.
%Our bound uses
%the monotone subsequences within each part (Lemma~\ref{opt:frontier-lemma}).
The correctness of this algorithm does
not depend on Theorem~\ref{thm:main}. Their other exact algorithm takes
$O(n\log^2 n)$ time; our bound is smaller, for example, when
$k=O(\log n)$. Both of our algorithms are comparison-based.

\subparagraph*{Related work.}
An \emph{$m$-modal} subsequence changes direction at most $m$ times.
Chung~\cite{Chung80} determined the largest length of such a subsequence
guaranteed in every $n$-point permutation for $m=1$, and conjectured
that this guarantee is asymptotic to $\sqrt{(2m+1)n}$ for fixed $m$.
Di~Stefano et al.~\cite{DiStefanoKLZ08} studied partitions into as few
$m$-modal subsequences as possible, proving NP-hardness for each
fixed $m\ge1$. Rollercoasters allow arbitrarily many direction changes
but impose a minimum run length, yielding a linear guarantee for
fixed $k$. Ahmed and Snevily~\cite{AhmedS13} studied permutations
that maximise the number of changes in direction.
They use the name ``roller coaster permutation''
for such a permutation, which is different from our definition
of a rollercoaster.

Adamson et al.~\cite{AdamsonFH24} allow weakly monotone
runs containing at least $k$ distinct values. For a word of length $n$
over $\sigma$ symbols, they compute the maximum length of such
rollercoasters in $O(n\sigma k)$ time and enumerate all maximum-length
ones with $O(n)$ delay after that preprocessing.
Fujita et al.~\cite{FujitaNIBT21} compute a longest common $k$-rollercoaster
of two integer strings of lengths $n_1,n_2$ in $O(n_1n_2k)$ time and
space.

\section{Definitions for sequences and rollercoasters}

We represent $S=(x_1,\ldots,x_n)$ by points $(a,x_a)$,
where we refer to the first item~$a$ as a \emph{position}
and the second item~$x_a$ as a \emph{value}.
A given point's
\emph{positional order} and \emph{value order}
are the numbers of other points with smaller position
and value than the given point, respectively.
A point~$p$ \emph{precedes}
another~$p'$ if $p$'s positional order is smaller than that of $p'$,
and lies \emph{above} or \emph{below} $p'$
if its value order is larger or smaller than those of $p'$, respectively. Write
$p\inc p'$ if $p$ precedes $p'$ and has smaller value, and $p\dec p'$
if it precedes $p'$ and has larger value.
A \emph{chain} is a monotone subsequence; its length is its number of points.
Let $\LIS(S)$ and $\LDS(S)$ denote the lengths of a longest increasing
and a longest decreasing subsequence of $S$, respectively, both zero
for the empty sequence.
The \emph{value complement} of a sequence replaces
each value $x$ by $n+1-x$, which flips the value order of points
while keeping their positional order.

A \emph{block} of a permutation is a set of consecutive positions whose
set of values is an interval of integers; the values within a block
may occur in any order. A permutation composed of blocks is therefore
specified by the number of points of each block, the relative order of
the values inside each block, and the order of the value intervals of
the blocks. For nonnegative integers $h,v$, let $\gridperm{h}{v}$ consist of
$h$ blocks, each a decreasing sequence of $v$ points, in increasing
value order. For example, $\gridperm{2}{3}=(3,2,1,6,5,4)$. When $h,v>0$, it
satisfies $\LIS(\gridperm{h}{v})=h$ and $\LDS(\gridperm{h}{v})=v$;
otherwise it is empty.

For $p\in S$, let $\ie{S}{p}$ and $\de{S}{p}$ be the lengths of
the longest increasing and decreasing chains ending at $p$,
respectively, and let $\is{S}{p}$ and $\ds{S}{p}$ be the corresponding
starting lengths. We call these four values the \emph{chain ranks} of
$p$ in $S$; the letters read increasing end, decreasing end,
increasing start, and decreasing start, respectively.
The subscript is essential: for a subsequence~$P$ of $S$
with $p\in P$,
$\ie{P}{p}$ is the length of the longest increasing chain
that ends at $p$ and only uses points in $P$.
Furthermore, extending the longest chains that end at~$p$
or start at $p'$ gives
\begin{equation}\label{eq:rank-order}
\begin{aligned}
 p\inc p'&\Longrightarrow \ie{S}{p'}>\ie{S}{p},\quad \is{S}{p}>\is{S}{p'},\\
 p\dec p'&\Longrightarrow \de{S}{p'}>\de{S}{p},\quad \ds{S}{p}>\ds{S}{p'}.
\end{aligned}
\end{equation}
Therefore, the pairs~$(\is{S}{p},\ds{S}{p})$
and $(\ie{S}{p},\de{S}{p})$ are unique with respect to the sequence~$S$.
We call the maps $p\mapsto(\ie{S}{p},\de{S}{p})$ and
$p\mapsto(\is{S}{p},\ds{S}{p})$ the \emph{ending rank-pair map} and the
\emph{starting rank-pair map} of $S$, respectively.
Joining two chains at their common endpoint also gives
\begin{equation}\label{eq:rank-sums}
 \ie{S}{p}+\is{S}{p}-1\le\LIS(S),\qquad
 \de{S}{p}+\ds{S}{p}-1\le\LDS(S).
\end{equation}
Finally, for nonnegative integers~$i$ and $j$,
let us define the sets
$\incclass{i}{S}=\{p:\ie{S}{p}=i\}$
and~$\decclass{j}{S}=\{p:\de{S}{p}=j\}$.
We use the following consequences of Seidenberg's argument.

\begin{lemma}[Seidenberg~\cite{Seidenberg59}]\label{lem:ranks}
The following are true.
\begin{enumerate}
  \item\label{lemitem:erdos} If $\LIS(S)\le x$ and $\LDS(S)\le y$ for nonnegative integers $x,y$,
  then $|S|\le xy$.
  \item\label{lemitem:incdecclassdef} The sets
  $\incclass{i}{S}$ and
  $\decclass{j}{S}$
  are decreasing and increasing chains, respectively. 
  \item\label{lemitem:incdecclassintersect} If $\LDS(S)\le q$,
  each decreasing chain $D$ of length~$q$ satisfies
  $|D\cap\decclass{j}{S}|=1$ for every $j\in[q]$.
  Symmetrically, if $\LIS(S)\le q$, each increasing chain $I$ of length~$q$
  satisfies $|I\cap\incclass{i}{S}|=1$ for every $i\in[q]$.
\end{enumerate}
\end{lemma}
\begin{proof}
The ending rank-pair map injectively maps each point in $S$
to a pair of integers in~$[x]\times [y]$,
and therefore $|S|\le xy$.
Further, two points of $\incclass{i}{S}$ cannot form an increasing pair, and
two points of $\decclass{j}{S}$ cannot form a decreasing pair.
Therefore $\incclass{i}{S}$ and $\decclass{j}{S}$
are decreasing and increasing chains, respectively.
Finally, if $\LDS(S)\le q$,
consider a decreasing $q$-point chain~$D$.
Each point in $D$ has a distinct $\de{S}$-rank
while its values are at most~$q$.
Therefore, $|D\cap\decclass{j}{S}|=1$ for every $j\in[q]$.
The proof for the $\LIS(S)\le q$ case is symmetric.
\end{proof}

For a subsequence~$P$ of $S$,
let us call the chain ranks~$\ie{S},\de{S},\is{S},\ds{S}$
for points in $P$ measured in $S$
as the \emph{ambient ranks} of each point.
The ambient ranks are important because some chain of $S$
involving some point~$p\in P$
may not be a subsequence of $P$.
For example, in $\gridperm{q}{2q+r}$ the middle $r$ points of each block form
the core. Since the ambient ranks satisfy $\de{S}{p}+\ds{S}{p}=2q+r+1$
for any $p\in P$,
we know the existence of a chain of length $2q+r$.
However, the chain ranks in $P$ alone only guarantee
a chain of length~$r$ from~$\de{P}{p}+\ds{P}{p}=r+1$.

\section{A sharp characterisation of \texorpdfstring{$\extremalf{k}{n}$}{f\_k(n)}}\label{sec:extremal}

Recall that $\extremalf{k}{n}$ is the minimum length of the longest
rollercoaster subsequence over all permutations of length $n$.
We now consider a dual function of $\extremalf{k}{n}$
that indicates the length of the longest permutation
whose longest rollercoaster subsequence is
at most a given length~$L$.
Formally, let $\extremalN{k}{L}=\max\{|S|:\rclen{k}{S}\le L\}$
be the extremal function for integers~$L\ge0$.
The function is nondecreasing and well-defined
since $\LIS(S),\LDS(S)\le\max(q,L)$
and Lemma~\ref{lem:ranks} bounds $|S|$ by $\max(q,L)^2$.
We prove the duality in Lemma~\ref{lem:duality}.

\begin{lemma}\label{lem:duality}
For all integers $n,L\ge0$,
$\extremalf{k}{n}\le L$ if and only if $n\le \extremalN{k}{L}$.
\end{lemma}
\begin{proof}
If $\extremalf{k}{n}\le L$,
there exists an $n$-point permutation $S$ with $\rclen{k}{S}\le L$,
so $n\le \extremalN{k}{L}$.
Conversely, if $n\le \extremalN{k}{L}$,
then there exists a sequence~$S$ with at least $n$ points
with~$\rclen{k}{S}\le L$.
We remove points in $S$ to obtain
an $n$-point subsequence~$S'$.
Then, we have~$\extremalf{k}{n}\le \rclen{k}{S'}\le \rclen{k}{S}\le L$,
where the first inequality is from the definition
of $\extremalf{k}{n}$
and the second from the fact that
a rollercoaster in $S'$ is also
a rollercoaster in $S$.
\end{proof}

Our strategy is to prove the equivalent form of Theorem~\ref{thm:main},
which is Theorem~\ref{thm:size}.
We prove the cases $L\le2q$ in Section~\ref{sec:short}, including
permutations of the maximum size, and the case $L>2q$ in
Section~\ref{sec:long}.

\begin{theorem}\label{thm:size}
For $k\ge3$, $q=k-1$, and every integer $L\ge0$,
\[
 \extremalN{k}{L}=
 \begin{cases}
 q^2,&0\le L\le q,\\
 2q^2-(2q-L)^2,&q<L<2q,\\
 qL,&L\ge2q.
 \end{cases}
\]
The middle expression agrees with the other two at $L=q$ and $L=2q$.
\end{theorem}

\subparagraph*{Where a rollercoaster can end or start.}
Towards counting the total number of points
in a sequence whose longest rollercoaster has length~$L$,
we identify the points at which no rollercoaster ends and
those at which none starts.
Let~$\rcend{p}$ and $\rcstart{p}$ be the lengths of the longest rollercoaster
ending and starting at $p$, or zero if no such rollercoaster exists.
Let~$\Zend=\{p:\rcend{p}=0\}$ and $\Zstart=\{p:\rcstart{p}=0\}$.
Note that these sets are not necessarily disjoint.
A rollercoaster ends at $p$ exactly when a chain of at least $q+1$
points ends there: the last run of a rollercoaster is a chain
with at least $q+1$ points, and conversely the chain itself is a rollercoaster. The same
argument applies to starting points. Hence
\[
 \rcend{p}>0\Longleftrightarrow\max\{\ie{S}{p},\de{S}{p}\}>q,
 \qquad
 \rcstart{p}>0\Longleftrightarrow\max\{\is{S}{p},\ds{S}{p}\}>q.
\]
In other words,
points in $\Zstart$ and $\Zend$
respectively have their starting and ending rank-pair in $[q]^2$.
Therefore, we have
$|\Zend|,|\Zstart|\le q^2$.
On the other hand, for points~$p\notin\Zend\cup\Zstart$,
we can join a longest rollercoaster ending at $p$
to a longest rollercoaster starting at $p$.
Therefore we have the following.

\begin{proposition}\label{prop:joiningCoasters}
  If $p\notin\Zend\cup\Zstart$, then
  $\rcend{p}+\rcstart{p}-1\le\rclen{k}{S}$.
\end{proposition}

\subsection{The case for short rollercoasters}\label{sec:short}
\subparagraph*{The range $0\le L\le q$.}
Here no rollercoaster is permitted, since its length would be at least~$q+1$.
Every chain therefore has at most $q$ points, giving
$\extremalN{k}{L}\le q^2$. Equality holds for $\gridperm{q}{q}$, which contains no chain
of $q+1$ points.

\subparagraph*{The upper-bound for $q\le L\le2q$.}
We prove that $|S|\le2q^2-(2q-L)^2$ holds
whenever we have~$\rclen{k}{S}\le L$.
First, since any increasing subsequence
of length at least $q+1$ is a rollercoaster,
we have $\LIS(S),\LDS(S)\le L$.
No point can be both an end and a start of a rollercoaster, since
joining a rollercoaster ending there to one starting there would give
at least $2q+1$ points, contradicting Proposition~\ref{prop:joiningCoasters}.
Therefore we have $S=\Zend\cup\Zstart$.

We now count the points in $\Zend\cup\Zstart$.
Since~${|\Zstart|\le q^2}$,
it remains to count points in $\Zend\setminus \Zstart$.
Each point~$p$ in this set has an ending rank-pair in $[q]^2$.
Furthermore, $\ie{S}{p}$ and $\de{S}{p}$ cannot be at least $L-q+1$
at the same time: otherwise the inequalities
in~\eqref{eq:rank-sums} would then give $\is{S}{p}$ and $\ds{S}{p}\le q$,
which gives $p\in\Zstart$. We thus exclude $(2q-L)^2$ pairs of
possible ending ranks, which finally yields
$\extremalN{k}{L}\le2q^2-(2q-L)^2$.

\subparagraph*{The matching construction.}
Let $t=2q-L$ and $u=q-t$. We construct a sequence
with~$q^2$ points
in each of $\Zend$ and $\Zstart$, and exactly $t^2$ points in their
intersection.
Since $|\Zend|=|\Zstart|=q^2$, we consider
a $q$-by-$q$ grid where we place the points in $\Zend$ and $\Zstart$
respectively by their ending and starting rank-pair maps.
Then, the points that map to $[u+1,q]\times [u+1,q]$
are exactly the points in $\Zend\cap\Zstart$.
Extending $u$ along the axes of the grid,
we partition the points in $\Zend$ and $\Zstart$
respectively into four subsets,
depending on whether or not each chain rank 
exceeds $u$.
In total, this gives us a partition
of $\Zend\cup \Zstart$ into seven subsets.

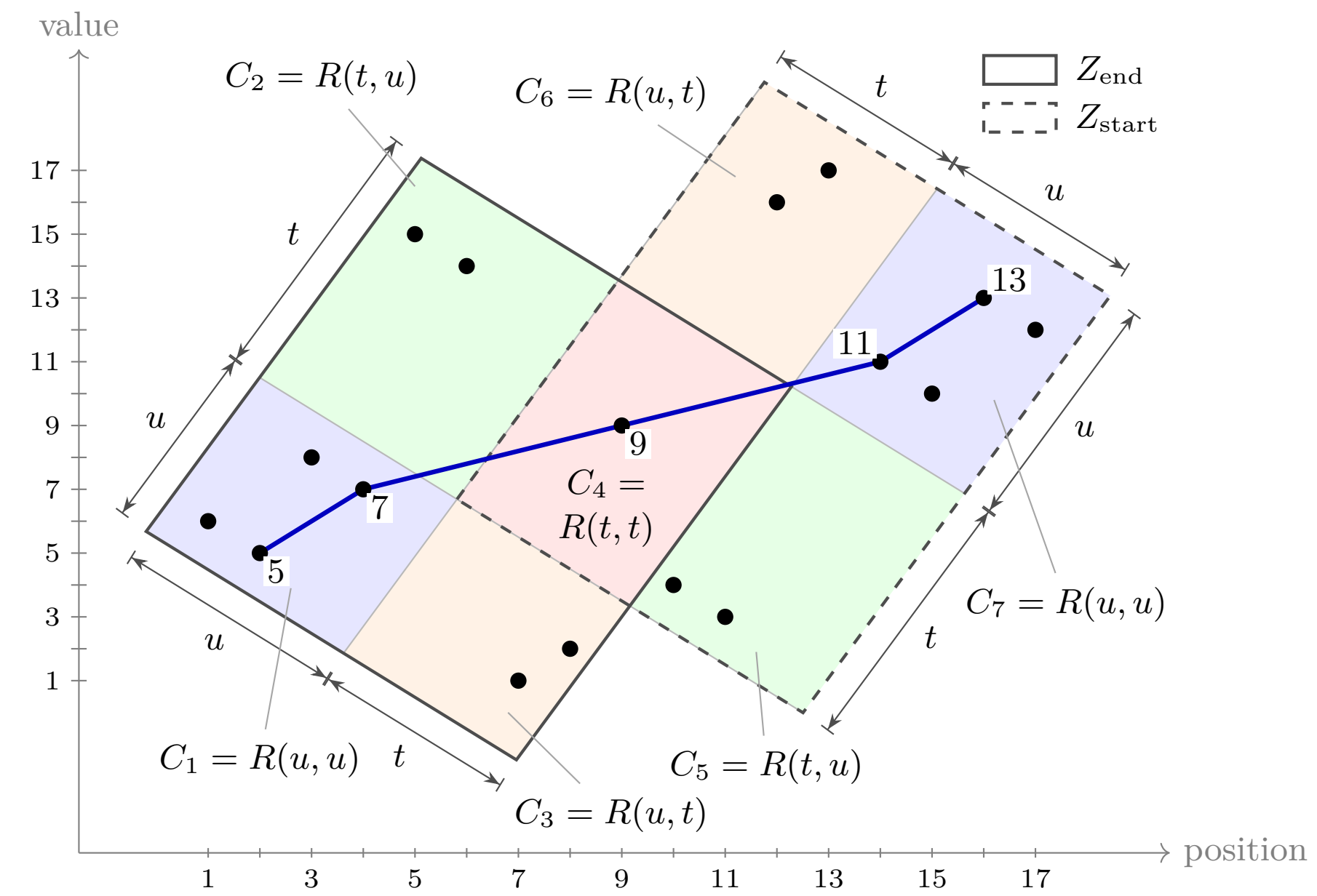
\begin{figure}[tb]
\centering
\begin{tikzpicture}[x=.47cm,y=.29cm,every node/.style={font=\small},
 width arrow/.style={black!70,thin,{Bar[width=4pt]Stealth[length=4pt]}-{Stealth[length=4pt]Bar[width=4pt]}},
 leader/.style={gray!70,line width=0.4pt}]
 % the seven cells of the two sheared grids
 \fill[blue!10] (-0.199,5.678) -- (1.995,10.505) -- (5.812,6.688) -- (3.618,1.860) -- cycle;% $C_1$
 \fill[green!10] (1.995,10.505) -- (5.120,17.380) -- (8.938,13.562) -- (5.812,6.688) -- cycle;% $C_2$
 \fill[orange!10] (3.618,1.860) -- (5.812,6.688) -- (9.156,3.344) -- (6.962,-1.483) -- cycle;% $C_3$
 \fill[red!10] (5.812,6.688) -- (8.938,13.562) -- (12.281,10.219) -- (9.156,3.344) -- cycle;% $C_4$
 \fill[green!10] (9.156,3.344) -- (12.281,10.219) -- (15.630,6.870) -- (12.505,-0.005) -- cycle;% $C_5$
 \fill[orange!10] (8.938,13.562) -- (11.757,19.765) -- (15.100,16.421) -- (12.281,10.219) -- cycle;% $C_6$
 \fill[blue!10] (12.281,10.219) -- (15.100,16.421) -- (18.449,13.072) -- (15.630,6.870) -- cycle;% $C_7$
 % inner grid lines: constant increasing end rank (slope $-1$) and constant decreasing end rank (slope $2.2$)
 \draw[gray!55] (1.995,10.505) -- (12.505,-0.005);
 \draw[gray!55] (5.120,17.380) -- (15.630,6.870);
 \draw[gray!55] (3.618,1.860) -- (11.757,19.765);
 \draw[gray!55] (6.962,-1.483) -- (15.100,16.421);
 % $\Zend$ solid, $\Zstart$ dashed
 \draw[black!70,thick] (-0.199,5.678) -- (5.120,17.380) -- (12.281,10.219) -- (6.962,-1.483) -- cycle;
 \draw[black!70,thick,dashed] (5.812,6.688) -- (11.757,19.765) -- (18.449,13.072) -- (12.505,-0.005) -- cycle;
 % width arrows, 0.32 cm outside the outer sides of the two boxes, split at the grid lines
 \draw[width arrow] (-0.67,6.24) -- (1.53,11.06);   \node at (-0.01,9.17) {$u$};% left side of $\Zend$
 \draw[width arrow] (1.53,11.06) -- (4.65,17.94);   \node at (2.65,15.02) {$t$};
 \draw[width arrow] (6.66,-2.28) -- (3.31,1.06);    \node at (4.70,-1.36) {$t$};% lower left side of $\Zend$
 \draw[width arrow] (3.31,1.06) -- (-0.50,4.88);    \node at (1.12,2.22) {$u$};
 \draw[width arrow] (12.06,20.57) -- (15.41,17.22); \node at (14.02,19.64) {$t$};% upper right side of $\Zstart$
 \draw[width arrow] (15.41,17.22) -- (18.75,13.87); \node at (17.37,16.30) {$u$};
 \draw[width arrow] (18.92,12.51) -- (16.10,6.31);  \node at (17.95,8.89) {$u$};% right side of $\Zstart$
 \draw[width arrow] (16.10,6.31) -- (12.97,-0.56);  \node at (14.98,2.35) {$t$};
 % axes drawn along the edges of the canvas; ticks at the integer positions and values
 \draw[->,gray] (-1.5,-4.4)--(19.6,-4.4) node[right]{position};
 \draw[->,gray] (-1.5,-4.4)--(-1.5,20.8) node[above]{value};
 \foreach \v in {1,...,17}{
  \draw[gray] (-1.65,\v)--(-1.35,\v);
  \draw[gray] (\v,-4.55)--(\v,-4.25);
 }
 \foreach \v in {1,3,...,17}{
  \node[left,font=\scriptsize] at (-1.60,\v) {$\v$};
  \node[below,font=\scriptsize] at (\v,-4.50) {$\v$};
 }
 \draw[blue!75!black,very thick] (2,5)--(4,7)--(9,9)--(14,11)--(16,13);
 \foreach \x/\y in {1/6,2/5,3/8,4/7,5/15,6/14,7/1,8/2,9/9,
  10/4,11/3,12/16,13/17,14/11,15/10,16/13,17/12}{
  \fill (\x,\y) circle[radius=2.1pt];
 }
 \foreach \x/\y in {2/5,4/7,9/9}{
  \node[below right=1pt,font=\small,fill=white,inner sep=1pt] at (\x,\y) {$\y$};
 }
 \node[above left=1pt,font=\small,fill=white,inner sep=1pt] at (14,11) {$11$};
 \node[above right=1pt,font=\small,fill=white,inner sep=1pt] at (16,13) {$13$};
 % cell names and internal permutations, placed outside the boxes and joined to a point inside
 % their cell by a thin leader line; C3 and C5 sit below the boxes, where no arrow runs
 \node (c1) at (2.00,-1.50)  {$C_1=\gridperm{u}{u}$};
 \node (c2) at (3.20,19.90)  {$C_2=\gridperm{t}{u}$};
 \node (c3) at (8.80,-3.20)  {$C_3=\gridperm{u}{t}$};
 \node[align=center,inner sep=1pt] at (8.7,6.4) {$C_4=$\\$\gridperm{t}{t}$};
 \node (c5) at (11.80,-1.70) {$C_5=\gridperm{t}{u}$};
 \node (c6) at (8.80,19.40)  {$C_6=\gridperm{u}{t}$};
 \node (c7) at (17.60,3.40)  {$C_7=\gridperm{u}{u}$};
 \draw[leader] (c1) -- (2.6,3.9);
 \draw[leader] (c2) -- (5.0,16.5);
 \draw[leader] (c3) -- (6.8,0.0);
 \draw[leader] (c5) -- (11.6,1.9);
 \draw[leader] (c6) -- (11.2,16.8);
 \draw[leader] (c7) -- (16.2,9.8);
 % legend for the two sets
 \draw[black!70,thick] (16.0,19.7) rectangle (17.4,20.6);
 \node[right] at (17.5,20.15) {$\Zend$};
 \draw[black!70,thick,dashed] (16.0,18.2) rectangle (17.4,19.1);
 \node[right] at (17.5,18.65) {$\Zstart$};
\end{tikzpicture}
\caption{The permutation $\sevenperm{3}{1}$, with $q=3$, $t=1$, $u=2$.
The solid box $\Zend$ and the
dashed box $\Zstart$ each form a skewed $q$-by-$q$ grid
whose points correspond to the ending and starting
chain ranks, respectively.
The two boxes share the central $t$-by-$t$ grid~$C_4$.
The highlighted rollercoaster
$(5,7,9,11,13)$ has length $2u+t=5$.}
\label{fig:block-template}
\end{figure}

We realise each subset using blocks~$\gridperm{h}{v}$,
where $h$ and $v$ are the heights and widths of the seven sub-grids
obtained by splitting $[q]^2$.
Specifically, consider the blocks $\gridperm{u}{u}$, $\gridperm{t}{u}$, $\gridperm{u}{t}$, $\gridperm{t}{t}$,
$\gridperm{t}{u}$, $\gridperm{u}{t}$, $\gridperm{u}{u}$,
each named $C_1,\ldots, C_7$ in order.
We assemble these blocks into a permutation~$\sevenperm{q}{t}$ so that the
blocks in increasing positional order are in the order of increasing subscripts,
and the blocks in increasing value order are $C_3,C_5,C_1,C_4,C_7,C_2,C_6$.
Figure~\ref{fig:block-template} draws
$\sevenperm{3}{1}$, with $q=3$, $t=1$, and $u=2$.

Any increasing or decreasing subsequence
in the first and last four blocks of $\sevenperm{q}{t}$
cannot be of length $q+1$.
Therefore, the points respectively in the first and last four blocks
are exactly the points in $\Zend$ and $\Zstart$.
Moreover, every point belongs
to at least one of $\Zend$ and $\Zstart$,
so no rollercoaster subsequence of $\sevenperm{q}{t}$ can have two or more runs.
Thus, the longest chain has length $u+t+u=2q-t$; one chain of this
length passes through $C_1,C_4,C_7$. Thus $|\sevenperm{q}{t}|=2q^2-t^2$ and
$\rclen{k}{\sevenperm{q}{t}}=2q-t$ for $t<q$, while $\rclen{k}{\sevenperm{q}{q}}=0$.
We prove these claims in Appendix~\ref{sec:construction}, including
how $\sevenperm{q}{t}$ avoids a longer chain.

\subsection{The case for long rollercoasters}\label{sec:long}

We first prove that $\extremalN{k}{L}$
is no greater than $\extremalN{k}{L-q}+q^2$
whenever $L\ge 2q$.
Then, for $L=2q+r+mq\ge 2q$,
where $m\ge0$ and $0\le r<q$,
we can repeatedly apply the bound
to obtain $\extremalN{k}{L}\le \extremalN{k}{2q+r}+mq^2$.

\begin{lemma}\label{lem:window}
For every $L\ge2q$, $\extremalN{k}{L}\le \extremalN{k}{L-q}+q^2$.
\end{lemma}
\begin{proof}
Let $S$ be a sequence satisfying $\rclen{k}{S}\le L$.
Let $H=\{p:\rcend{p}>L-q\}$,
and assume towards contradiction that there exists
a $(q+1)$-point chain in $H$.
Then, from the point with the smallest positional order
in this $(q+1)$-point chain,
we extend the rollercoaster ending at this point
with the chosen chain.
This gives us a rollercoaster of at least $L+1$ points,
which contradicts $\rclen{k}{S}\le L$.
Thus, we have $\LIS(H),\LDS(H)\le q$, and $|H|\le q^2$ by
Lemma~\ref{lem:ranks}.
Furthermore, every rollercoaster in $S\setminus H$
has length at most its original $\rcend{p}$,
which is at most $L-q$.
\end{proof}

We now prove $\extremalN{k}{2q+r}\le2q^2+qr$,
which will give us $\extremalN{k}{L}\le q(2q+r+mq)=qL$.
For a sequence~$S$ with $\rclen{k}{S}\le2q+r$,
let us define the \emph{core} of $S$
as~$\core{S}=S\setminus (\Zstart\cup \Zend)$.
Then, every point in the core is both an end and a start
of some rollercoaster, and $|S|\le 2q^2+|\core{S}|$.
Thus, it suffices to prove that $|\core{S}|\le qr$,
which we will do in the remainder of this section.

\begin{lemma}\label{lem:core-new}
  For $r<q=k-1$, a permutation~$S$
  with $\rclen{k}{S}\le 2q+r$
  and point~$p\in \core{S}$,
  \begin{flalign*}
    &\rcend{p}=\max(\ie{S}{p},\de{S}{p})>q\text{ and }
    \rcstart{p}=\max(\is{S}{p},\ds{S}{p})>q,\\
    &\max(\ie{S}{p},\de{S}{p})+\max(\is{S}{p},\ds{S}{p})\le2q+r+1,\\
    &\LIS(\core{S}),\LDS(\core{S})\le q \text{ and }\ie{S}{p},\de{S}{p},\is{S}{p},\ds{S}{p}
    \in [q+r].
  \end{flalign*}
  In particular, $\core{S}=\emptyset$ when $r=0$.
\end{lemma}
\begin{proof}
For every $p\in \core{S}$,
we have $\rcend{p},\rcstart{p}\ge q+1$
while Proposition~\ref{prop:joiningCoasters} gives
$\rcend{p}+\rcstart{p}\le2q+r+1$. However, this is impossible for $r=0$
and therefore $\core{S}$ is empty.
Otherwise, we have $\rcend{p},\rcstart{p}\le q+r<2q+1$,
thus a longest rollercoaster ending at $p$ and a
longest one starting at $p$ each have only one run.
Furthermore, assume towards contradiction
that $\core{S}$ contains a $(q+1)$-point chain.
Joining this chain
to a rollercoaster ending at its first point and one starting at its
last point would give a rollercoaster of at least $3(q+1)-2=3q+1>2q+r$ points,
which contradicts $\rclen{k}{S}\le 2q+r$.
This proves that $\LIS(\core{S}),\LDS(\core{S})\le q$.
\end{proof}

\subparagraph*{Restructuring a core for easy counting.}\label{sec:layers}
We first introduce a property on permutations
based on whether it can be represented in a specific form.
This representation has three advantages.
First, we can isolate the points in $\core{S}$ from $S$
and parameterise the values~$q$ and~$r$
while selectively remembering the chain ranks of each point in $S$
as labels.
Second, we can partition the points in such a permutation
by the equivalence of labels, which allows us to bound
the total number of points in terms of $r$.
Finally, any subsequence of a permutation with this property
also has this property, which enables induction on the parameter~$q$.

\begin{definition}\label{def:layers}
Let $c,w$ be strictly positive integers. A permutation $P$ \emph{admits
a \layerrep{c}{w}} if there exist a partition of $P$
into subsequences $U,M,D$ and labels $\estrand(p),\sstrand(p)\in[w]$ at
every point $p$ such that:
{\upshape (i)} every value in $U$ exceeds every value in $M\cup D$, and every
value in $D$ is smaller than every value in $U\cup M$;
{\upshape (ii)} $\LIS(P),\LDS(P)\le c$, and $\estrand(p)+\sstrand(p)\le w+1$ for
every $p\in P$;
{\upshape (iii)} for pairs $p,p'$ in positional order, we have:

\begin{tabular}{c|c|c|c}
 &layers containing $p,p'$&comparison&labels\\\hline
 \strandc{1}&$U\cup M$&$p\inc p'$&$\estrand(p)<\estrand(p')$\\
 \strandc{2}&$M\cup D$&$p\inc p'$&$\sstrand(p)>\sstrand(p')$\\
 \strandc{3}&$D$&$p\dec p'$&$\estrand(p)<\estrand(p')$\\
 \strandc{4}&$U$&$p\dec p'$&$\sstrand(p)>\sstrand(p')$.
\end{tabular}

We call \strandc{1} to \strandc{4}
\emph{strand constraints},
and such a partition of $P$, together with the labels at each point,
is called a \emph{\layerrep{c}{w}} of $P$.
\end{definition}

For a point~$p$ in a permutation admitting a \layerrep{c}{w},
we call the values~$\estrand(p)$ and $\sstrand(p)$
respectively the \emph{ending and starting strand} of $p$.
Note that any layer may be empty, and the layers
need not be consecutive in positional order.
We now prove that $\core{S}$ or its value complement has this property
in Lemma~\ref{lem:core-representation}.

\begin{lemma}\label{lem:core-representation}
For $r<q=k-1$ and a permutation~$S$
with $\rclen{k}{S}\le 2q+r$, $\core{S}$ or its value
complement admits a \layerrep{q}{r}.
\end{lemma}
\begin{proof}[Proof sketch]
Every point~$p\in \core{S}$ with $\ie{S}{p}>q$ and $\de{S}{p}\le q$
lies above every point~$p'\in \core{S}$ with
$\de{S}{p'}>q$ and $\ie{S}{p'}\le q$.
Assume otherwise towards contradiction;
then $p\inc p'$ contradicts $\ie{S}{p}>q$ and $\ie{S}{p'}\le q$,
and $p'\dec p$ contradicts $\de{S}{p}\le q$ and $\de{S}{p'}>q$.
Therefore we can cut $\core{S}$ horizontally
so that points~$p$ above and below the cut
respectively have $\ie{S}{p}>q$ and $\de{S}{p}>q$.
Starting ranks similarly admit a cut selecting
$\ds{S}{p}>q$ above and $\is{S}{p}>q$ below.
If the first cut is above the second,
we complement the values in~$\core{S}$.
Let $U, M$ and $D$
be the set of points above both cuts,
between cuts and below both cuts, respectively.
We choose $\estrand(p)=\ie{S}{p}-q$, $\sstrand(p)=\ds{S}{p}-q$ for $p\in U$,
$\estrand(p)=\ie{S}{p}-q$, $\sstrand(p)=\is{S}{p}-q$ for $p\in M$, and
$\estrand(p)=\de{S}{p}-q$, $\sstrand(p)=\is{S}{p}-q$ for $p\in D$.
This choice of strands satisfies the strand constraints
and $\estrand(p)+\sstrand(p)\le r+1$ for all $p$.
\end{proof}

Consider the points in the middle layer~$M$
in a \layerrep{c}{w} of a permutation~$P$.
The ending strand increases whenever the value of a point increases,
and since $\estrand(p)+\sstrand(p)\le w+1$,
we have $\LIS(M)\le w$.
Furthermore, we can bound $\LIS(P)+\LIS(M)$
by exchanging prefixes between
longest increasing subsequences in $P$ and $M$.

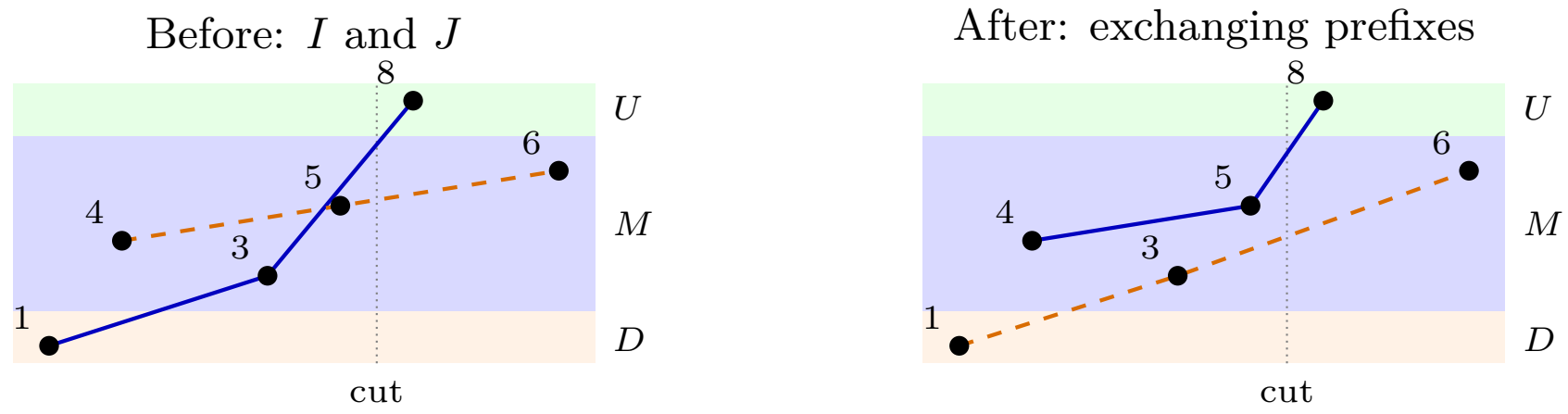
\begin{figure}[htbp]
\centering
\begin{tikzpicture}[x=.52cm,y=.25cm,every node/.style={font=\small}]
 \foreach \shift in {0,12.5}{
  \begin{scope}[xshift=\shift*.52cm]
   \fill[orange!10] (.5,.5) rectangle (8.5,2);
   \fill[blue!15] (.5,2) rectangle (8.5,7);
   \fill[green!10] (.5,7) rectangle (8.5,8.5);
   \draw[densely dotted,gray] (5.5,.5)--(5.5,8.5);
   \node[below,font=\scriptsize] at (5.5,.5) {cut};
   \node[right,font=\scriptsize] at (8.5,1.2) {$D$};
   \node[right,font=\scriptsize] at (8.5,4.5) {$M$};
   \node[right,font=\scriptsize] at (8.5,7.8) {$U$};
  \end{scope}
 }
 \draw[blue!75!black,thick] (1,1)--(4,3)--(6,8);
 \draw[orange!85!black,thick,dashed] (2,4)--(5,5)--(8,6);
 \foreach \x/\y in {1/1,4/3,6/8,2/4,5/5,8/6}{
  \fill (\x,\y) circle[radius=2pt];
  \node[above left,font=\scriptsize] at (\x,\y) {$\y$};
 }
 \node[above] at (4.5,9.0) {Before: $I$ and $J$};
 \begin{scope}[xshift=6.5cm]
  \draw[blue!75!black,thick] (2,4)--(5,5)--(6,8);
  \draw[orange!85!black,thick,dashed] (1,1)--(4,3)--(8,6);
  \foreach \x/\y in {1/1,4/3,6/8,2/4,5/5,8/6}{
   \fill (\x,\y) circle[radius=2pt];
   \node[above left,font=\scriptsize] at (\x,\y) {$\y$};
  }
  \node[above] at (4.5,9.0) {After: exchanging prefixes};
 \end{scope}
\end{tikzpicture}
\caption{Exchanging prefixes just before the point with value $8$ replaces
$(1,3,8)$ and $(4,5,6)$ by $(1,3,6)$ and $(4,5,8)$. The new chains
avoid $U$ and $D$, respectively; point labels are values.}
\label{fig:uncross}
\end{figure}

\begin{lemma}\label{lem:uncross}
If $P$ admits a \layerrep{c}{w} with layers
$U,M,D$, then we have
$\LIS(P)+\LIS(M)\le\LIS(U\cup M)+\LIS(M\cup D)\le2w$.
\end{lemma}
\begin{proof}[Proof sketch]
Take longest increasing chains $I$ in $P$ and $J$ in $M$.
We produce an increasing chain in $M\cup D$ and an increasing chain in
$U\cup M$ whose lengths sum to $|I|+|J|$. \textbf{Case 1:} $I$ avoids~$U$
or avoids $D$. Then $I$ and $J$ themselves are the two
chains. \textbf{Case 2:} $J$ is empty. Then $M$ is empty and thus
we split $I$ into two sequences so that each new sequence
is entirely in $U$ and $D$.
\textbf{Case 3:} $I$ and $J$ share a point $p\in M$.
We exchange their parts
after $p$: both new chains contain $p$, so the lengths still sum to
$|I|+|J|$; the chain beginning with the part of $I$ before $p$ avoids
$U$, and the other avoids $D$. 
\textbf{Case 4:} $I$ and $J$ are
disjoint and $I$ meets both outer layers. We cut both chains before a
point $x$ of $I$ chosen so that the exchanged prefixes join
increasingly, as in Figure~\ref{fig:uncross}, where $x$ is the point
of value $8$; again the chain containing the prefix of $I$ avoids $U$
and the other avoids $D$.
This proves the first inequality.
The definition of a \layerrep{c}{w}
restricts the length of increasing chains
in $U\cup M$ and $M\cup D$ by
$w$, thus the second inequality follows.
\end{proof}

Let us now replace the labels
such that $\estrand(p)=\ie{M}{p}$ and $\sstrand(p)=\is{M}{p}$
for all $p\in M$.
Note that this labelling scheme greedily maps
each point to a strand at the earliest point possible,
and thus any label satisfying
the definition of a \layerrep{c}{w}
has labels no less than the new labels.
Thus, we have
$1\le \estrand(p),\sstrand(p)\le \LIS(M)$,
$\estrand(p)+\sstrand(p)\le \LIS(M)+1$
and $\estrand(p)+\sstrand(p)\le w+1$
for the new labels.
Furthermore, such a relabelling preserves the comparison results
between labels between a pair of points,
and thus the resulting representation
is also a \layerrep{c}{w}.
Let us call the new labels the \emph{tightened labels}
of $M$.

Let us partition the points in $U$ and $D$
by their ending and starting strands, respectively.
Let $U_i=\{p\in U:\estrand(p)=i\}$
and $D_j=\{p\in D:\sstrand(p)=j\}$
be the respective sets in each partition.
Then, the sets $U_i$ and $D_j$ are decreasing. Along $U_i$, the starting
strands decrease strictly within $[w+1-i]$; along $D_j$, the ending
strands increase strictly within $[w+1-j]$. Consequently,
for the points in $U$ and $D$, we have
$|U_i|\le w+1-i$ and $|D_j|\le w+1-j$
for all $1\le i,j\le w$.
The next lemma restricts the position of points in $M$
with respect to the points
in the upper and lower sets.
This also restricts the decreasing ranks of points in $M$.

\begin{lemma}\label{lem:precedence}
For $p\in M$, every point of $U_{i'}$ with $i'\le\estrand(p)$ precedes
$p$, and every point of $D_{j'}$ with $j'\le\sstrand(p)$ follows $p$.
\end{lemma}
\begin{proof}
Suppose, for a contradiction, that some point~$p'$ of $U_{i'}$ with
$i'\le\estrand(p)$ follows $p$. Since its value is larger,
this contradicts the strand constraint \strandc{1}.
Similarly, if a point of $D_{j'}$ with
$j'\le\sstrand(p)$ precedes $p$,
this again contradicts \strandc{2}.
\end{proof}

From Lemma~\ref{lem:precedence}, we now bound the number
of different rank-pairs
the points in $M$ can map to with respect to the sequence~$P$.
Let $\Ustar{i}$ be a largest of the sets $U_1,\ldots,U_i$ and
$\Dstar{j}$ a largest of $D_1,\ldots,D_j$.
Attach $\Ustar{\estrand(p)}$ before $p$,
or $\Dstar{\sstrand(p)}$ after $p$. Joining decreasing chains at~$p$ also gives
$\de{P}{p}\ge|\Ustar{\estrand(p)}|+1$,
$\ds{P}{p}\ge|\Dstar{\sstrand(p)}|+1$
and $\de{P}{p}+\ds{P}{p}\le c+1$.
Both maps $p\mapsto(\estrand(p),\de{P}{p})$
and $p\mapsto(\sstrand(p),\ds{P}{p})$
are injective on $M$,
since for $p,p'\in M$, $p\inc p'$ changes the labels by the strand constraints,
and $p\dec p'$ changes the decreasing ranks.
Let us choose a parameter~$\tau$ with $0\le\tau\le c$.
We count points in $M$ with $\de{P}\le\tau$ by the
first map.
Then, the remaining points have $\ds{P}\le c-\tau$; we count them
by the second. Letting $(z)_+=\max(z,0)$, this proves Lemma~\ref{lem:split}.
Furthermore, we bound each half of this count in Lemma~\ref{lem:triangle}.

\begin{lemma}\label{lem:split}
After tightening labels, for $h=\LIS(M)$ and every integer
$0\le\tau\le c$,
\begin{equation}\label{eq:split}
 |P|\le
 \sum_{i=1}^w |U_i|+\sum_{i=1}^h(\tau-|\Ustar{i}|)_+
 +\sum_{j=1}^w |D_j|+\sum_{j=1}^h(c-\tau-|\Dstar{j}|)_+.
\end{equation}
\end{lemma}

\begin{lemma}\label{lem:triangle}
Let $w,h$ be integers with $w\ge1$ and $0\le h\le w$, and let
$U_1,\ldots,U_w$ be finite sets with $|U_i|\le w+1-i$ for $i\in[w]$.
Let $\Ustar{i}$ be a largest of $U_1,\ldots,U_i$. If the nonnegative integer
$\tau$ satisfies $2\tau+h\le2w+1$, then
\begin{equation}\label{eq:triangle}
 \sum_{i=1}^w |U_i|+\sum_{i=1}^h(\tau-|\Ustar{i}|)_+
 \le\frac{w(w+1)}2.
\end{equation}
\end{lemma}
\begin{proof}
Let $m$ be the maximum strand with $m\le h$ and $|\Ustar{m}|<\tau$,
with $m=0$ if no such $m$ exists.
Consider adding the $i$th term of each summation.
For $i\le m$, we have $|U_i|+(\tau-|\Ustar{i}|)\le\tau$.
Since there are $m$ terms and $m\le h$,
the first $m$ terms sum up to $m\tau\le m(2w+1-m)/2$.
For $i>m$, we have $\tau-|\Ustar{i}|\le 0$ and thus
the second summation contributes 0. Overall, the sum is at most
\begin{equation*}
\frac{m(2w+1-m)}{2}+\sum_{i=m+1}^w|U_i|\le
\frac{m(2w+1-m)}{2}+\sum_{i=m+1}^w(w+1-i)\le
 \frac{w(w+1)}{2}.\qedhere
\end{equation*}
\end{proof}

\begin{proposition}\label{prop:joint}
If $P$ admits a \layerrep{c}{w} whose middle layer
$M$ satisfies $c+\LIS(M)\le2w$, then $|P|\le w(w+1)$.
\end{proposition}
\begin{proof}
Choose $\tau=\floor{c/2}$ in~\eqref{eq:split}. Since
$2\ceil{c/2}+h\le c+1+h\le2w+1$, both $\tau$ and $c-\tau$ satisfy
Lemma~\ref{lem:triangle}. Apply it to the sets $U_i$ with cutoff $\tau$,
and to the sets $D_j$ with cutoff $c-\tau$
using $|U_i|\le w+1-i$ and $|D_j|\le w+1-j$.
Each half of~\eqref{eq:split} is at most $w(w+1)/2$.
\end{proof}

\subparagraph*{An example of the combined count.}
Let $c=4$, $w=3$, and $P=(7,1,4,3,6,5,8,2)$, with
$U=(7,8)$, $M=(4,3,6,5)$, and $D=(1,2)$.
Assign labels $(\estrand,\sstrand)=(1,3),(3,1)$ to the points of each
outer layer in positional order, and $(1,2),(1,2),(2,1),(2,1)$ to $M$.
With these labels, $P$ admits a \layerrep{4}{3};
here $\LIS(P)=\LDS(P)=4$, $h=2$, and $c+h=2w$.
Here $(|U_1|,|U_2|,|U_3|)=(|D_1|,|D_2|,|D_3|)=(1,0,1)$ and
$|\Ustar{1}|=|\Ustar{2}|=|\Dstar{1}|=|\Dstar{2}|=1$.
With $\tau=2$, each half of~\eqref{eq:split} is $2 + (2 - 1) + (2 - 1) = 4$,
and this gives $8$ points in total. Lemma~\ref{lem:triangle} bounds each half by
$3+2+1=6$, yielding $|P|\le12=cw$.

\begin{theorem}\label{thm:capacity}
For strictly positive integers $c,w$, the largest size of a permutation~$P$
that admits a \layerrep{c}{w} 
is $c\min(c,w)$.
\end{theorem}
\begin{proof}
We prove by induction on $c$
with induction hypothesis $|P|\le c\min(c,w)$.
We have two base cases and an induction.

\textbf{Base case 1:} if $c\le w$,
then we have $\LIS(P),\LDS(P)\le c\le w$.
Then, by item~\ref{lemitem:erdos} of Lemma~\ref{lem:ranks},
we have $|P|\le c^2=c\min (c,w)$.

\textbf{Base case 2:} if $c>w$ and $c\le \LIS(P)$,
let $M$ be the middle layer
of such a \layerrep{c}{w}.
Then, we have $c+\LIS(M)\le \LIS(P)+\LIS(M)\le 2w$,
where the second inequality is from Lemma~\ref{lem:uncross}.
Proposition~\ref{prop:joint} then gives $|P|\le w(w+1)\le c\min(c,w)$.

\textbf{Induction:} we have $c>w$ and $c>\LIS(P)$.
From the definition of a \layerrep{c}{w},
we also have $c\ge \LDS(P)$.
Recall the partition $P$ into $c$ sets
$\decclass{j}{P}$
of increasing points from item~\ref{lemitem:incdecclassdef}
of Lemma~\ref{lem:ranks}.
Choose one of the smallest sets,
and delete the set from $P$.
If the chosen set is empty,
no decreasing chain of length $c$ existed in~$P$.
Otherwise,
every decreasing chain of length $c$ uses some point in this set
by item~\ref{lemitem:incdecclassintersect} of Lemma~\ref{lem:ranks}.
Thus, the remaining permutation $Q$ satisfies $\LIS(Q),\LDS(Q)\le c-1$.
Note that the remaining points admit a \layerrep{c-1}{w}.
Since we delete at most $|P|/c$ points,
the induction hypothesis gives $\frac{c-1}{c}|P|\le |Q|\le(c-1)w$.
In other words, we have $|P|\le cw=c\min (c,w)$.
\end{proof}

We apply Lemma~\ref{lem:core-representation}
to plug in $\core{S}$ or its value complement
in Theorem~\ref{thm:capacity}.
This gives the upper-bound in the following theorem.

\begin{theorem}\label{thm:core-size}
  For $r<q=k-1$
  and a permutation~$S$ with $\rclen{k}{S}\le 2q+r$,
  we have $|\core{S}|\le q\min(q,r)$, and this bound is sharp.
\end{theorem}
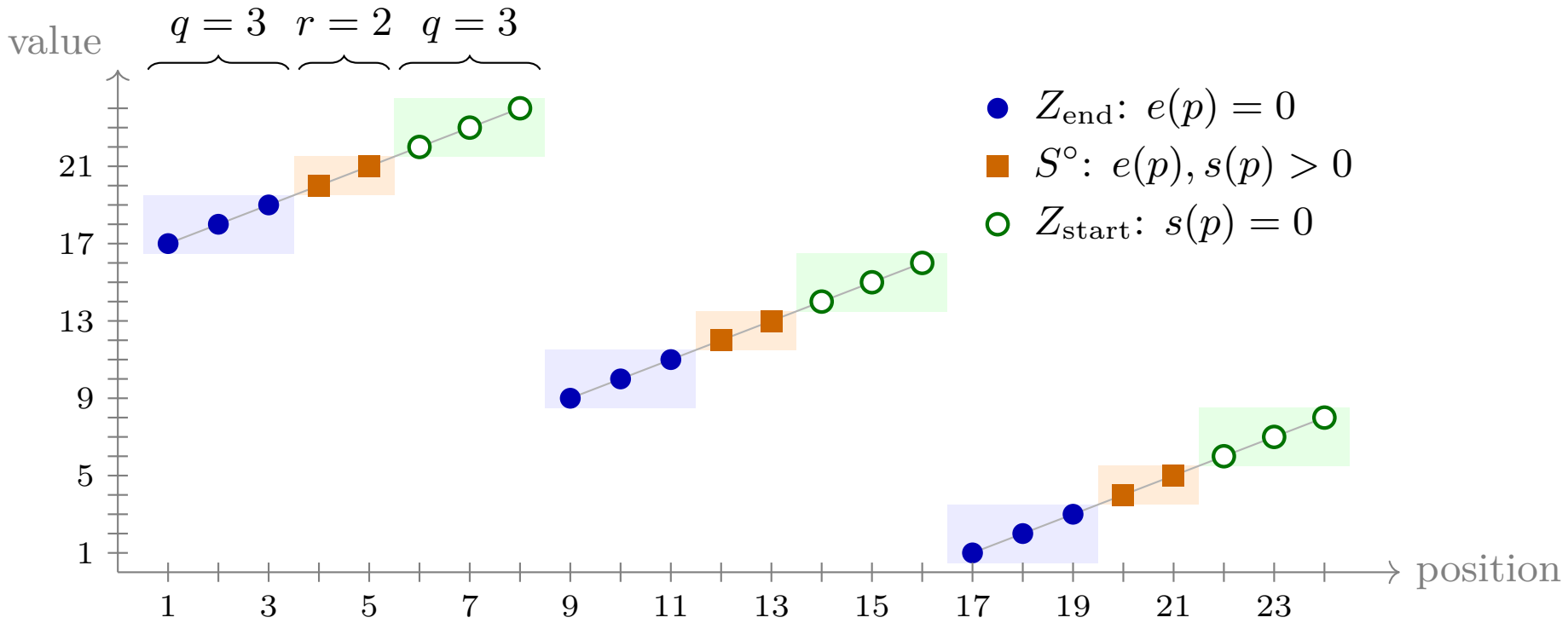
\begin{figure}[htbp]
\centering
\begin{tikzpicture}[x=.39cm,y=.15cm,every node/.style={font=\small}]
 \foreach \offset/\base in {0/16,8/8,16/0}{
  \fill[blue!8] ({\offset+.5},{\base+.5}) rectangle ({\offset+3.5},{\base+3.5});
  \fill[orange!15] ({\offset+3.5},{\base+3.5}) rectangle ({\offset+5.5},{\base+5.5});
  \fill[green!10] ({\offset+5.5},{\base+5.5}) rectangle ({\offset+8.5},{\base+8.5});
  \draw[gray!60] ({\offset+1},{\base+1})--({\offset+8},{\base+8});
  \foreach \ell in {1,2,3}{
   \fill[blue!70!black] ({\offset+\ell},{\base+\ell}) circle[radius=2.3pt];
  }
  \foreach \ell in {4,5}{
   \path[fill=orange!80!black] ({\offset+\ell},{\base+\ell})
    +(-2.4pt,-2.4pt) rectangle +(2.4pt,2.4pt);
  }
  \foreach \ell in {6,7,8}{
   \draw[green!45!black,fill=white,thick] ({\offset+\ell},{\base+\ell}) circle[radius=2.3pt];
  }
 }
 \draw[->,gray] (0,0)--(25.5,0) node[right]{position};
 \draw[->,gray] (0,0)--(0,26) node[above left]{value};
 \foreach \v in {1,...,24}{
  \draw[gray] (\v,-.5)--(\v,.5);
  \draw[gray] (-.2,\v)--(.2,\v);
 }
 \foreach \v in {1,3,...,23}{\node[below,font=\scriptsize] at (\v,-.4) {$\v$};}
 \foreach \v in {1,5,...,21}{\node[left,font=\scriptsize] at (-.15,\v) {$\v$};}
 \draw[decorate,decoration={brace,amplitude=3pt}] (.6,26)--(3.4,26)
  node[midway,above=3pt,text height=1.5ex,text depth=.25ex] {$q=3$};
 \draw[decorate,decoration={brace,amplitude=3pt}] (3.6,26)--(5.4,26)
  node[midway,above=3pt,text height=1.5ex,text depth=.25ex] {$r=2$};
 \draw[decorate,decoration={brace,amplitude=3pt}] (5.6,26)--(8.4,26)
  node[midway,above=3pt,text height=1.5ex,text depth=.25ex] {$q=3$};
 \fill[blue!70!black] (17.5,24) circle[radius=2.3pt];
 \node[right] at (17.9,24) {$\Zend$: $\rcend{p}=0$};
 \path[fill=orange!80!black] (17.5,21)
    +(-2.4pt,-2.4pt) rectangle +(2.4pt,2.4pt);
 \node[right] at (17.9,21) {$\core{S}$: $\rcend{p},\rcstart{p}>0$};
 \draw[green!45!black,fill=white,thick] (17.5,18) circle[radius=2.3pt];
 \node[right] at (17.9,18) {$\Zstart$: $\rcstart{p}=0$};
\end{tikzpicture}
\caption{The points of $\gridperm{3}{8}$ in reverse positional order, with
$q=3$, $r=2$: the six squares form the core and induce $\gridperm{3}{2}$ in
reverse positional order; the three points before and after them in
their block add $3$ to both of their increasing ranks.}
\label{fig:core-plot-main}
\end{figure}

For the sharpness,
we construct a permutation~$\gridperm{q}{2q+\min(q,r)}$
with $q\min(q,r)$ points
in Figure~\ref{fig:core-plot-main}.
Summing up,
we have $\extremalN{k}{L}\le \extremalN{k}{2q+r}+mq^2\le qL$
from Lemma~\ref{lem:window} and Theorem~\ref{thm:core-size}.
Furthermore, consider the permutation~$\gridperm{q}{L}$;
any decreasing chain is a subsequence of a single block,
and an increasing chain uses at most one point per block.
Thus, every rollercoaster consists of a single decreasing chain
of length at most $L$, giving
$\extremalN{k}{L}\ge qL$.
Therefore we have $\extremalN{k}{L}=qL$ for the $L\ge 2q$ case.
Combined with the length bound for short rollercoasters,
we have Theorem~\ref{thm:size};
Lemma~\ref{lem:duality} then gives Theorem~\ref{thm:main}.

\section{Algorithms for rollercoaster subsequences}\label{sec:algorithms}
In addition to our combinatorial bounds,
we outline two more algorithms for computing rollercoaster subsequences
of a given sequence, provided that each value is distinct and is over
a totally ordered set.
Since only the relative order matters,
we treat the sequence~$S=(x_1,\ldots,x_n)$ as a permutation
over the set of elements. We work in the standard word-RAM model with word
size $\Omega(\log n)$ and constant-time comparisons between elements.
The output is a sequence of positions that designate a subsequence of $S$.

We first describe an algorithm that computes
a rollercoaster subsequence with at least $\extremalf{k}{n}$ points.
If $q^2<n\le2q^2$, we compute each of the four chain ranks
whose values are capped at $2q$ through four scans over $S$.
In each scan, we also leave pointers that allow us to
backtrack and retrieve
the longest increasing or decreasing subsequence
starting and ending at each point,
or such a subsequence of length at least $2q$.
By Theorem~\ref{thm:main},
there exists a rollercoaster subsequence
consisting of a single run of length $\extremalf{k}{n}$,
or two runs each with $q+1$ points,
either of which we can compute from the four chain ranks at each point.

For $n>2q^2$, we first identify shorter rollercoasters
and then extend these by~$q$ points every step. Choose $L=\ceil{n/q}=\beta+Tq$
to be the length of the rollercoaster subsequence we seek,
where $\beta,T$ are integers with $2q+1\le\beta\le3q$ and $T\ge0$.
We mark every point at which a rollercoaster of at least
$\beta$ points ends while maintaining pointers that allow backtracking
to retrieve such a rollercoaster. We find these rollercoasters by
joining at most three
monotone subsequences of at least $q+1$ points at shared endpoints
(Lemma~\ref{lem:seeds}).

After the marking,
initialise empty sets~$\class{0},\class{1},\ldots,\class{T}$
called \emph{classes}.
Then, for each marked point~$p$ in increasing positional order,
put $p$ into a class~$\class{h}$,
where $h$ is some positive integer
such that $\class{h-1}\cup\{p\}$ has a monotone $(q+1)$-point subsequence
and $\class{h}\cup\{p\}$ does not.
Let $h=0$ if no such integer exists.
This choice of $h$ satisfies
the following invariants after putting each point into
each chosen class:
no class contains a monotone $(q+1)$-point subsequence,
and each point in $\class{h}$ is the end of a rollercoaster of at least
$\beta+hq$ points.
There are more than $Tq^2$ marked points in $S$;
the set of points without marks
does not have a rollercoaster of length~$\beta$,
and therefore has at most $\extremalN{k}{\beta-1}=q(\beta-1)< |S|-Tq^2$ points.
Since each class has at most $q^2$ points by the first invariant
and Lemma~\ref{lem:ranks},
some point~$p$ will be put in $\class{T}$.
Then, the second invariant gives a rollercoaster of length at least $L=\beta+Tq$.
We describe the details of the algorithm and prove its correctness in
Appendix~\ref{app:algorithm}.

\begin{theorem}\label{thm:algorithm}
Given a sequence of $n>(k-1)^2$ distinct elements of a totally
ordered set, we can compute and return a $k$-rollercoaster subsequence
of length at least $\extremalf{k}{n}$
in~$O(n\log \extremalf{k}{n})$ time and $O(n)$ space
using only comparisons between elements.
\end{theorem}

Next, we improve the $O(nk^2)$-time algorithm of
Gawrychowski et al.~\cite{GawrychowskiMS22} for the problem
of computing a longest rollercoaster.
First, in the preprocessing phase,
their algorithm partitions $S$ into minimal contiguous sequences
that contain a monotonic subsequence of length $k$,
which we call \emph{segments}, and then partitions
each segment into at most $k$
monotone subsequences in $O(n\log k)$ time.
Then, for each element~$x_j$, the main algorithm
computes the length of the longest subsequence ending at $x_j$
that is also a rollercoaster up to the second-to-last run;
the last run in the subsequence may have fewer than $k$ points.
We call such a sequence a \emph{pseudocoaster}.
This is achieved by dynamic programming
where each subproblem is defined on the ending position~$j$,
the length of the final run ending at $j$, capped at $k$,
and whether the final run was increasing or decreasing,
and we process the positions in increasing order.
Each subproblem investigates points to the left of $j$
and extends the longest pseudocoaster ending there by a single point~$j$,
if possible.
A key observation by Gawrychowski et al.~\cite{GawrychowskiMS22}
localises the occurrence of any second-to-last point~$p$
in a longest pseudocoaster ending at $j$.
Specifically, $p$ lies in the segment
including $j$ or one of the preceding four. The subproblem can then be evaluated
depending on the subproblems for only the most recent five segments.

We accelerate the search for $p$ with $2(k-1)$ search structures, one
for each direction and the length $s\le k$
of a pseudocoaster's final run.
Each search structure maintains 
a subset of points in the preceding four segments
and the points visited in the current segment,
along with the length of the longest pseudocoaster ending at each point
whose final run has the given direction and length as weights.
The selected points correspond to critical values where the weights change,
and are sorted in increasing order with respect to their weights.
At each point $j$,
we make one query to each structure that asks for the point~$p$
associated with the greatest weight
among points with $x_p<x_j$ for an increasing run,
or $x_p>x_j$ for a decreasing run.
We add one to these values and take the maximum over them
to complete the evaluation of each subproblem,
and insert $j$ associated with the computed value
into the corresponding search structure.

The key to efficiency is to maintain $k$ pointers in the search structures,
one for each monotone subsequence of the segment,
each of which remembers the output for the previous query
in each subsequence.
Since the subsequences are monotone, each pointer only increases or decreases
depending on the direction of the run.
We use (2,4) trees that are augmented with
pointers called \emph{fingers}~\cite{HuddlestonM82} as the base data structure,
which allows queries and insertions in $O(\log k)$ amortised
time each
at the cost of linear initialisation time
and amortised constant bookkeeping time. 
We give a detailed description of the algorithm
as well as its correctness and running-time proof in Appendix~\ref{opt:app:algorithm}.
Note that the following result is independent
of Theorem~\ref{thm:main}.

\begin{theorem}\label{opt:thm:longest}
Given $n\ge1$ distinct elements of a totally ordered set and an integer
$k\ge3$, we can compute and return a longest $k$-rollercoaster in
$O(nk\log k)$ time and $O(nk)$ space, using only comparisons between
elements. If no $k$-rollercoaster
exists, the algorithm reports an empty subsequence.
\end{theorem}

\section{Further questions}\label{sec:conclusion}
For the problem of finding a rollercoaster
with at least $f_k(n)$ points,
our algorithm runs in $O(n\log \extremalf{k}{n})=O(n\log (n/q))$ time,
which is $O(n\log k)$ for $n\in O(k^2)$.
It is open whether or not we can solve the same problem in $O(n\log k)$ time
regardless of $n$ and $k$,
or in $O(n)$ time independently of $k$.
For the problem of computing
a longest rollercoaster, Theorem~\ref{opt:thm:longest}
gives $O(nk\log k)$ time and 
Gawrychowski et al.~\cite{GawrychowskiMS22} give $O(n\log^2 n)$
against their lower bound of $\Omega(n\log k)$ comparisons. This gap remains open.

%% file: appendix.tex
\section*{Technical appendix}
\begin{figure}[htbp]
\centering
\begin{tikzpicture}[every node/.style={font=\small},
 box/.style={draw=black!55,rectangle split,rectangle split parts=2,
             align=center,inner sep=4pt},
 wide/.style={box,text width=11.2cm},
 narrow/.style={box,text width=3.2cm},
 grey/.style={rectangle split part fill={gray!22,gray!8}},
 blue/.style={rectangle split part fill={blue!18,blue!7}},
 orange/.style={rectangle split part fill={orange!25,orange!10}},
 green/.style={rectangle split part fill={green!20,green!8}},
 decision/.style={draw=black!55,fill=gray!8,diamond,aspect=2.6,
                  minimum width=4.4cm,minimum height=1.5cm,inner sep=1pt,align=center},
 arr/.style={->,thick,draw=black!65,rounded corners=3pt},
 lab/.style={font=\small,align=center,inner sep=1.5pt,fill=white}]
 % the quantity to bound
 \node[wide,grey,anchor=north] (start) at (0,0)
   {\textbf{Goal}
    \nodepart{two}
    Bound $\extremalN{k}{L}$, the largest size of a permutation in which
    every rollercoaster has at most $L$ points};
 % decision on the range of L
 \node[decision,below=5mm of start] (d1) {range of $L$};
 \draw[arr] (start.south) -- (d1.north);
 \node[narrow,blue,anchor=north] (short) at ([yshift=-5mm]-4.1,0 |- d1.south)
   {\textbf{Section~\ref{sec:short}}
    \nodepart{two}
    Count points where rollercoasters cannot end or start};
 \node[narrow,orange,anchor=north] (mid) at ([yshift=-5mm]0,0 |- d1.south)
   {\textbf{Section~\ref{sec:long}}
    \nodepart{two}
    Write $L=2q+r$ with $0\le r<q$};
 \node[narrow,orange,anchor=north] (long) at ([yshift=-5mm]4.1,0 |- d1.south)
   {\textbf{Lemma~\ref{lem:window}}
    \nodepart{two}
    Count and remove at most $q^2$ points};
 \draw[arr] (d1.west) -| (short.north) node[lab,pos=0.25,above] {$L\le2q$};
 \draw[arr] (d1.south) -- (mid.north) node[lab,midway,right=1pt] {$2q\le L<3q$};
 \draw[arr] (d1.east) -| (long.north) node[lab,pos=0.25,above] {$L\ge3q$};
 % loop of the long range: back to the start with L-q
 \draw[arr] (long.east) -- ++(4mm,0) |- ([yshift=4mm]d1.north east) -- (d1.north east);
 \node[lab,right=1pt] at ([xshift=4mm]long.east |- d1.south) {repeat\\with\\$L-q$};
 % the core
 \node[draw=black!55,fill=orange!10,align=center,inner sep=4pt,text width=8cm,anchor=north]
   (core) at ([yshift=-7mm]0,0 |- long.south)
   {At most $2q^2$ points are not both an end and a start of a rollercoaster;
    the other points form the core $\core{S}$};
 \draw[arr] (mid.south) -- (core.north);
 % layers and the theorem
 \node[wide,green,below=5mm of core] (layers)
   {\textbf{Lemma~\ref{lem:core-representation}}
    \nodepart{two}
    Reformulate $\core{S}$ into a \layerrep{q}{r}};
 \draw[arr] (core.south) -- (layers.north);
 % decision on the case of the proof, directly below the layers
 \node[decision,below=5mm of layers] (d2) {};
 \node[align=center] at (d2.center) {A $(c,w)$-representation};
 \draw[arr] (layers.south) -- (d2.north);
 \node[narrow,blue,anchor=north] (base1) at ([yshift=-5mm]-4.1,0 |- d2.south)
   {\textbf{Lemma~\ref{lem:ranks}, Item~\ref{lemitem:erdos}}
    \nodepart{two}
    $\LIS,\LDS\le c$ gives $c^2$ points};
 \node[narrow,blue,anchor=north] (base2) at ([yshift=-5mm]0,0 |- d2.south)
   {\textbf{Proposition~\ref{prop:joint}}
    \nodepart{two}
    Split the representation using thresholds $\floor{c/2}$
    and $\ceil{c/2}$,
    count $w(w+1)/2$ points in each area};
 \node[narrow,blue,anchor=north] (induct) at ([yshift=-5mm]4.1,0 |- d2.south)
   {\textbf{Lemma~\ref{lem:ranks}, Item~\ref{lemitem:incdecclassintersect}}
    \nodepart{two}
    Count and remove
    at most $n/c$ points
    from a smallest class of equal decreasing end rank};
 \draw[arr] (d2.west) -| (base1.north) node[lab,pos=0.25,above] {$c\le w$};
 \draw[arr] (d2.south) -- (base2.north) node[lab,font=\footnotesize,inner sep=.5pt,midway,right=2pt] {$c>w$, $\LIS(P)=c$};
 \draw[arr] (d2.east) -| (induct.north) node[lab,font=\footnotesize,pos=0.30,above] {$c>w,\LIS(P)<c$};
 % loop of the induction: back to the case decision with c-1
 \draw[arr] (induct.east) -- ++(4mm,0) |- ([yshift=4mm]d2.north east) -- (d2.north east);
 \node[lab,right=1pt] at ([xshift=4mm]induct.east |- d2.south) {repeat\\with\\$c-1$};
 % the theorem at the bottom, reached from both base cases
 \node[wide,green,anchor=north] (capacity) at ([yshift=-5mm]0,0 |- induct.south)
   {\textbf{Theorem~\ref{thm:capacity}}
    \nodepart{two}
    A permutation with a \layerrep{c}{w} has at most $c\min(c,w)$ points};
 \draw[arr] (base1.south) -- (capacity.north -| base1.south);
 \draw[arr] (base2.south) -- (capacity.north -| base2.south);
\end{tikzpicture}
\caption{The structure of the proof for Theorem~\ref{thm:size}.}
\label{fig:roadmap}
\end{figure}
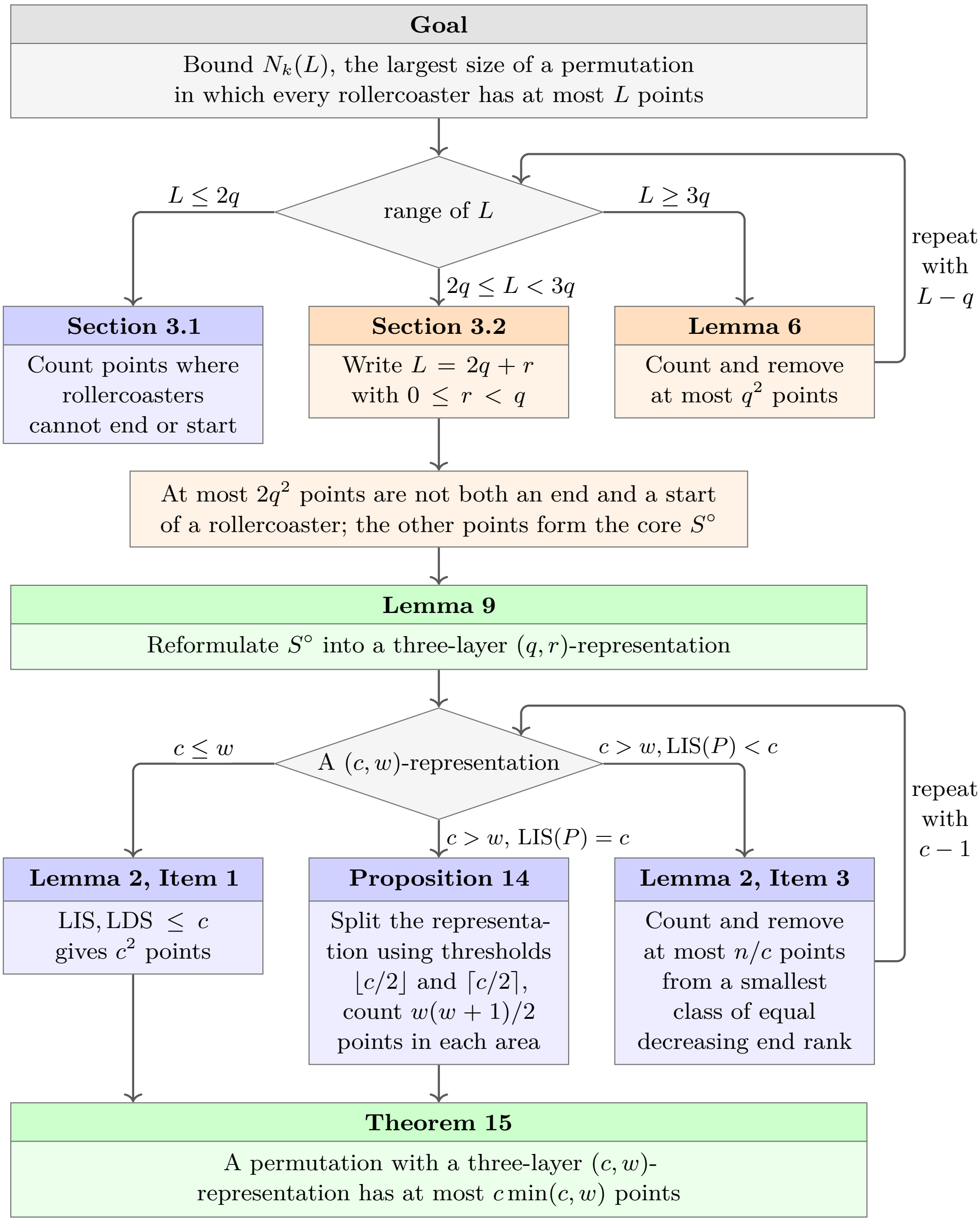

The appendices contain the proofs omitted from
Sections~\ref{sec:extremal} and \ref{sec:algorithms}.
Specifically,
in Appendix~\ref{app:constructions},
we describe the construction for $\sevenperm{q}{t}$
in detail,
which proves the tightness of $\extremalN{k}{L}=2q^2-(2q-L)^2$
for the $q\le L\le 2q$ regime.
In Appendix~\ref{app:capacity},
we give detailed proofs for Lemmas~\ref{lem:core-representation}
and~\ref{lem:uncross}.
We also provide examples for counting points in
a \layerrep{c}{w} using Lemma~\ref{lem:split}.
We conclude the proof for Theorem~\ref{thm:main}
in Appendix~\ref{app:inversion}.
In Appendices~\ref{app:algorithm} and \ref{opt:app:algorithm},
we describe and prove the correctness and running times
of the two algorithms,
respectively proving Theorems~\ref{thm:algorithm} and~\ref{opt:thm:longest}.
We summarise the combinatorial argument
in Figure~\ref{fig:roadmap},
where the proof for the upper-bound is in
Section~\ref{sec:extremal}.
\section{The matching construction for the short rollercoaster case}\label{app:constructions}\label{sec:construction}
For $L$ with $q\le L\le 2q$,
we prove that the number of points in the seven-block construction~$\sevenperm{q}{t}$
matches the upper-bound obtained in Section~\ref{sec:short}.
Let $t=2q-L$, $u=q-t$ and~$W=\sevenperm{q}{t}$.
Then, we have $0\le t\le q$
and $L=2q-t=2u+t$.
Recall that $\gridperm{h}{v}$ consists of~$h$ blocks in increasing value order,
where each block consists of $v$ decreasing points.
When~$h,v>0$, the~$\ell$th point in the $\nu$th block of $\gridperm{h}{v}$
has ending rank-pair~$(\nu,\ell)$ and starting rank-pair~$(h+1-\nu,v+1-\ell)$.
Thus the images of both rank-pair maps of $\gridperm{h}{v}$
are exactly~$[h]\times[v]$.
The permutation is empty if either parameter is zero.

Our goal is to construct a sequence
with $q^2$ points in each of $\Zend$ and
$\Zstart$ and $t^2$ points in their intersection.
The strategy is to divide the $q$-by-$q$ image of
each rank-pair map into four rectangles~$[u]\times [u]$,
$[u+1,q]\times [u]$, $[u]\times [u+1,q]$
and $[u+1,q]\times [u+1,q]$.
Note that the final rectangle is a square that has exactly $t^2$ points.
We choose this square to be the intersection
between $\Zstart$ and $\Zend$,
which can be achieved
by the seven blocks~$C_1,\ldots,C_7$ appearing in
the order specified in Table~\ref{tab:construction}.

\begin{table}[htbp]
\centering\small
\begin{tabular}{@{}lccccccc@{}}
\toprule
Block in positional order &$C_1$&$C_2$&$C_3$&$C_4$&$C_5$&$C_6$&$C_7$\\
\midrule
Value order, lowest first &3&6&1&4&2&7&5\\
Internal permutation &$\gridperm{u}{u}$&$\gridperm{t}{u}$&$\gridperm{u}{t}$&$\gridperm{t}{t}$&$\gridperm{t}{u}$&$\gridperm{u}{t}$&$\gridperm{u}{u}$\\
\bottomrule
\end{tabular}
\caption{The seven blocks that constitute $W$.}
\label{tab:construction}
\end{table}

Let $\eta=\max(2u,q)$.
We prove bounds for ending chain ranks of points in $W$
in Lemma~\ref{lem:construction-ending}.
We can repeat a symmetric argument from right to left
to obtain bounds for starting chain ranks
in Corollary~\ref{cor:construction-starting}.

\begin{lemma}\label{lem:construction-ending}
  For a point~$p\in W$,
  the values~$\ie{W}(p)$ and $\de{W}(p)$
  are no greater than the following values,
  depending on where $p$ is from.
  
  \begin{center}
    \begin{tabular}{rrrrrrrr}
    \toprule
    $p$ is from & $C_1$&$C_2$&$C_3$&$C_4$&$C_5$&$C_6$&$C_7$\\
    \midrule
    $\ie{W}(p)\le$&$u$&$q$&$u$&$q$&$q$&$L$&$L$\\
    $\de{W}(p)\le$&$u$&$u$&$q$&$q$&$L$&$t$&$\eta$\\
    \bottomrule
    \end{tabular}
  \end{center}
\end{lemma}
\begin{proof}
  We prove the statements in the increasing order of
  $i=1,\ldots,7$.

  \textbf{Case 1:} $p\in C_1$. Since $C_1$ is the leftmost block,
  we have $\ie{W}(p),\de{W}(p)\le u\le q$
  from~$\LIS(\gridperm{u}{u})=\LDS(\gridperm{u}{u})=u$.
  
  \textbf{Case 2:}  $p\in C_2$.
  We have $\de{W}(p)\le u$
  since $C_1$ is below $C_2$
  and $\LDS(\gridperm{t}{u})=u$.
  Furthermore, we have
  $\ie{W}(p)\le \LIS(C_1)+\LIS(C_2)= u+t= q$.

  \textbf{Case 3:} $p\in C_3$.
  We have $\ie{W}(p)\le u$
  since $C_1$ and $C_2$ are above $C_3$
  and $\LIS(\gridperm{u}{t})=u$.
  Furthermore,
  any decreasing subsequence
  ending at a point in~$C_1$ or~$C_2$
  cannot use points from~$C_2$ or~$C_1$,
  respectively.
  Therefore, we have
  $\de{W}(p)\le \max(\LDS(C_1),\LDS(C_2)) + \LDS(C_3)
  =u+t=q$.

  \textbf{Case 4:} $p\in C_4$.
  Any increasing subsequence
  ending at a point in $C_1$ or $C_3$
  cannot use points from $C_3$ or $C_1$,
  respectively.
  Since $C_2$ is above $C_4$
  and $C_1$ and $C_3$ are below $C_4$, we have $\ie{W}(p)\le
  \max(\LIS(C_1),\LIS(C_3)) + \LIS(C_4)
  =u+t=q$.
  Likewise, we have $\de{W}(p)
  \le \LDS(C_2)+\LDS(C_4)=u+t=q$.

  \textbf{Case 5:} $p\in C_5$.
  Note that $C_3$ is below $C_5$,
  and $C_1$, $C_2$ and $C_4$ are above $C_5$.
  For any decreasing chain ending at $p$,
  let $p'$ be the rightmost point in the chain
  not in $C_5$.
  Then, we have $\de{W}(p)\le \de{W}(p')+\LDS(C_5)
  =\max(u,u,q) + u=q+u=L$.
  Furthermore,
  we have $\ie{W}(p)\le \LIS(C_3)+\LIS(C_5)=u+t=q$.

  \textbf{Case 6:} $p\in C_6$.
  Note that $C_1,\ldots,C_5$
  are all below $C_6$.
  Therefore, $\de{W}(p)\le \LDS(C_6)=t$.
  Likewise, for any increasing chain
  ending at $p$,
  let $p'$ be the rightmost point in the chain
  not in $C_6$.
  Then, we have $\ie{W}(p)\le \ie{W}(p')+\LIS(C_6)
  =\max(u,q,u,q,q) + u = q+u = L$.

  \textbf{Case 7:} $p\in C_7$.
  Then, $C_1,C_3,C_4, C_5$ are below $C_7$
  and $C_2, C_6$ are above $C_7$.
  By the same logic,
  we have $\ie{W}(p)\le \max(u,u,q,q) + \LIS(C_7)
  =q+u = L$,
  and $\de{W}(p)\le \max(u,t) + \LDS(C_7)
  =\max(u+u, u+t)=\max(2u,q)=\eta$.

  This completes the proof
  of the upper-bounds
  for $\ie{W}(p)$ and $\de{W}(p)$.
\end{proof}

\begin{corollary}\label{cor:construction-starting}
  For a point~$p\in W$,
  the values~$\is{W}(p)$ and $\ds{W}(p)$
  are no greater than the following values,
  depending on where $p$ is from.
  
  \begin{center}
    \begin{tabular}{rrrrrrrr}
    \toprule
    $p$ is from & $C_1$&$C_2$&$C_3$&$C_4$&$C_5$&$C_6$&$C_7$\\
    \midrule
    $\is{W}(p)\le$&$L$&$q$&$L$&$q$&$q$&$u$&$u$\\
    $\ds{W}(p)\le$&$\eta$&$L$&$t$&$q$&$u$&$q$&$u$\\
    \bottomrule
    \end{tabular}
  \end{center}
\end{corollary}

\begin{lemma}\label{lem:construction}
For $0\le t\le q$, the permutation $\sevenperm{q}{t}$ has $2q^2-t^2$ points,
contains no $k$-rollercoaster with two or more runs, and has maximum
monotone subsequence length $2q-t$. Consequently,
$\rclen{k}{\sevenperm{q}{t}}=L$ for $t<q$,
and $\rclen{k}{\sevenperm{q}{q}}=0$.
\end{lemma}
\begin{proof}
The number of points in the permutation
is~$|\sevenperm{q}{t}|=2u^2+4ut+t^2=2q^2-t^2$.
From Lemma~\ref{lem:construction-ending},
points~$p$ in the first four blocks of $W$ have
$\ie{W}(p),\de{W}(p)\le q$.
Thus, it follows that
$C_1\cup C_2\cup C_3\cup C_4\subseteq\Zend$.
Furthermore, we have $|C_1\cup C_2\cup C_3\cup C_4|=q^2$,
and since $\Zend$ can have at most $q^2$ points,
no other point can be in $\Zend$.
Likewise, from Corollary~\ref{cor:construction-starting},
we have $C_4\cup C_5\cup C_6\cup C_7\subseteq\Zstart$,
and by a symmetric argument we also achieve equality.
Therefore, we have $C_4=\Zend\cap \Zstart$,
which has exactly $t^2$ points by the definition
of $C_4=\gridperm{t}{t}$.
If a rollercoaster with at least two runs exists,
then two chains each of length at least $q+1$ must meet
at some point in $W$.
This contradicts $W=\Zend\cup \Zstart$, and therefore
no rollercoaster in $W$ has two or more runs.
Furthermore, from Lemma~\ref{lem:construction-ending}
and Corollary~\ref{cor:construction-starting},
the length of any rollercoaster is at most $L$.

For the matching lower bound, concatenate increasing chains of lengths
$u,t,u$ in $C_1,C_4,C_7$. These blocks increase in both position and
value, giving a chain of length $u+t+u=L$.
For the $t=0$ case,
$C_4$ is empty, and thus the chain has length $2q$.
Conversely, for the $t=q$ case,
$C_1$ and $C_7$ are empty. Then,
the longest increasing chain in $C_4$ has length $q$.
Therefore, we have $\rclen{k}{\sevenperm{q}{q}}=0$.
\end{proof}

\begin{example}\label{ex:seventeen}
Figure~\ref{fig:block-template}
describes $\sevenperm{q}{t}$
for $k=4$, $q=3$, $t=1$, and $u=2$.
The complete permutation is
$\sevenperm{3}{1}=(6,5,8,7\mid15,14\mid1,2\mid9\mid4,3\mid16,17\mid11,10,13,12)$,
where the separators mark $C_1,\ldots,C_7$.
Its longest $4$-rollercoaster has length $5$; one example is
$(5,7,9,11,13)$ through $C_1,C_4,C_7$, highlighted in
Figure~\ref{fig:block-template}, which draws this permutation with its
seven blocks. Here $C_4$ is the single point shared by $\Zend$
and $\Zstart$. Hence $f_4(17)=5$.
\end{example}

\FloatBarrier
\section{Full proofs and examples for Section~\ref{sec:long}}\label{app:capacity}

\subsection{Proof of Lemma~\ref{lem:core-representation}: the two value cuts}\label{app:cuts}
Let $S$ be a permutation with $\rclen{k}{S}\le2q+r$.
We first prove that the points~$x$ with $\ie{S}(x)>q$, $\de{S}(x)\le q$ lie above
the points~$y$ with $\de{S}(y)>q$, $\ie{S}(y)\le q$. Assume towards contradiction
that $x$ lies below $y$.
If $x$ precedes $y$, then $\ie{S}{y}>\ie{S}{x}>q$,
which contradicts $\ie{S}(y)\le q$.
If $y$ precedes $x$,
then $\de{S}{x}>\de{S}{y}>q$,
which contradicts $\de{S}(x)\le q$.
Therefore, any point~$x$ with $\ie{S}(x)>q$ and $\de{S}(x)\le q$
is above the points~$y$ with $\de{S}(y)>q$ and $\ie{S}(y)\le q$.

Let us choose a horizontal cut immediately above the highest point~$p$ with
$\ie{S}(p)\le q$, or below all points in $\core{S}$ if no such point exists.
Then, every point~$x$ above the cut has $\ie{S}(x)>q$.
Furthermore, a point~$y$ below the cut either
has both ending ranks greater than~$q$, or has $\de{S}(y)>q$ and
$\ie{S}(y)\le q$.
This is because
points~$x$ with only $\ie{S}(x)>q$ lie above every point~$y$ with
$\ie{S}(y)\le q$.
Thus, the chosen cut horizontally divides points
so that we have $\ie{S}>q$ above the cut and
$\de{S}>q$ below it.

Symmetrically, the points with $\ds{S}>q$, $\is{S}\le q$
lie above the points with $\is{S}>q$, $\ds{S}\le q$.
We therefore choose a second horizontal cut
so that points~$x$ above the cut satisfy $\ds{S}(x)>q$
and points~$y$ below the cut have $\is{S}(y)>q$.

If the first cut is above the second,
we complement the values of $S$ so that
the first cut is below the second.
Note that the core and rollercoasters in $S$ correspond to the core
and rollercoasters
of the value complement of $S$,
and thus no argument is lost in the complementation.
Let $U$ contain the points above both cuts, $M$ those between them,
and $D$ those below both cuts.
Note that if the cuts coincide, then we simply have $M=\emptyset$.
Since the cuts are horizontal,
every value in $U$ exceeds every value in $M\cup D$,
and every value in $D$ is smaller than that in $U\cup M$.

We choose $\estrand(p)=\ie{S}{p}-q$, $\sstrand(p)=\ds{S}{p}-q$ for $p\in U$,
$\estrand(p)=\ie{S}{p}-q$, $\sstrand(p)=\is{S}{p}-q$ for $p\in M$, and
$\estrand(p)=\de{S}{p}-q$, $\sstrand(p)=\is{S}{p}-q$ for $p\in D$.
The strands are then strictly positive and are in the form of
subtracting $q$ from some chain rank at $p$.
Thus, the $\max(\ie{S}{p},\de{S}{p})+\max(\is{S}{p},\ds{S}{p})\le2q+r+1$ condition
from Lemma~\ref{lem:core-new}
gives $\estrand+\sstrand\le r+1$.
Note that the strand constraints are all satisfied.
First, for any $p,p'\in U\cup M$ with $p\inc p'$,
we have $\estrand(p)=\ie{S}{p}-q < \ie{S}{p'}-q=\estrand(p')$.
Second, for $p,p'\in M\cup D$ with $p\inc p'$,
we have $\sstrand(p)=\is{S}{p}-q > \is{S}{p'}-q=\sstrand(p')$.
Third, for $p,p'\in D$ with $p\dec p'$,
we have $\estrand(p)=\de{S}{p}-q < \de{S}{p'}-q=\estrand(p')$.
Finally, for $p,p'\in U$ with $p\dec p'$,
we have $\sstrand(p)=\ds{S}{p}-q > \ds{S}{p'}-q=\sstrand(p')$.
Since we already have $\LIS(\core{S}), \LDS(\core{S})\le q$
from Lemma~\ref{lem:core-new},
the three layers and the chosen labels
satisfy the definition of a \layerrep{q}{r}.
This completes the proof of Lemma~\ref{lem:core-representation}.

\subsection{Sharpness in Theorem~\ref{thm:core-size}}\label{app:core-sharpness}
Let $m=\min(q,r)$ and $S=\gridperm{q}{2q+m}$. A decreasing chain of
$S$ lies in one block, and an increasing chain uses at most one point
per block, so every rollercoaster of $S$ is a decreasing chain inside
a block and $\rclen{k}{S}=2q+m\le2q+r$. At the $j$th point of a
block, $\de{S}=j$ and $\ds{S}=2q+m+1-j$, whereas $\ie{S},\is{S}\le q$.
Hence a rollercoaster ends at this point exactly when $j\ge q+1$, and
one starts there exactly when $j\le q+m$. The core $\core{S}$
therefore consists of the middle $m$ points of each block and has
$qm=q\min(q,r)$ points.
Figure~\ref{fig:core-plot-main} illustrates
this example in reverse positional order for $q=3$, $r=2$.

\subsection{Proof of Lemma~\ref{lem:uncross}: exchanging the chains}\label{app:exchange}
Take longest increasing chains $I$ in $P$ and $J$ in $M$.
We swap prefixes of $I$ and $J$ into increasing chains in $M\cup D$
and $U\cup M$.
If $J$ is empty, then $M$ is empty, and thus we split $I$ between
$D$ and $U$. If $I$ avoids an outer layer, $I$ and $J$ already
give the required chains. If $I,J$ share a point $p$, we exchange
their prefixes before $p$, keeping one copy of $p$ in
each chain. Both chains obtained this way are still increasing.
Since $p\in M$, the new
chain with the prefix of $I$ avoids $U$, and the other avoids $D$.

The remaining case is when $J$ and $I$ are disjoint,
but $J$ is not empty and $I$ has points in both $U$ and $D$.
For $t\in I$, let $z(t)$ be the last
point of $J$ preceding $t$, or the first point of $J$ if none
precedes $t$. Choose the first $x\in I$ that lies above $z(x)$.
The point $x$ exists because every point of $I\cap U$ lies above
$z(x)$, while no point of $I\cap D$ does.
Cut both chains immediately before $x$, into prefixes $I_-,J_-$
and suffixes $I_+,J_+$. Consider the new sequences~$I_-J_+$
and $J_-I_+$.
The second sequence is increasing: when $J_-$ is nonempty, its last
point is $z(x)$, below the first point $x$ of $I_+$.
For the first join, let $y$ be the last point of $I_-$.
The point~$y$ exists because there is a point in $I$ from $D$,
and the choice of $x$ places $y$ below $z(y)$.
When $J_+$ is nonempty, its first point $w$ is $z(y)$ or a later
point of the increasing chain $J$; therefore $y$ lies below $w$.
Finally, if $J_-$ or $J_+$ is empty,
the corresponding new sequence is $I_+$ or $I_-$, respectively,
and they are increasing.
Note that the prefix $I_-$ avoids $U$ by the
choice of $x$, and $I_+$ avoids $D$ because its first value is above
a value in $M$. 
Summing up, in every case, one new chain avoids $U$
while the other avoids $D$,
and the total length of the new chains is $|I|+|J|$.
This proves the first inequality.
Figure~\ref{fig:uncross} gives an example of exchanging prefixes of $I$ and $J$.

For the second inequality, \strandc{1} makes $\estrand$ strictly increasing on
an increasing chain in $U\cup M$, and \strandc{2} makes $\sstrand$ strictly
decreasing on one in $M\cup D$. Both labels are in $[w]$, so each
chain has at most $w$ points.

\subsection{Examples for Lemma~\ref{lem:split} and Proposition~\ref{prop:joint}}\label{app:count-example}
We provide examples for counting
the number of points in a \layerrep{c}{w}
using Proposition~\ref{prop:joint}.
We identify each summand appearing in Lemma~\ref{lem:split},
bound the number of points
in each half of the representation in Lemma~\ref{lem:triangle}, and
combine these numbers to obtain the count
in Proposition~\ref{prop:joint}.

\begin{figure}[!htbp]
\centering
\begin{tikzpicture}[x=.31cm,y=.29cm,every node/.style={font=\small},
 threshold/.style={red!65!black,line width=1.1pt},
 dim/.style={black,line width=.6pt,{Bar[width=4pt]Stealth[length=4pt]}-{Stealth[length=4pt]Bar[width=4pt]}},
 mk/.style={fill=blue!40!black,inner sep=0pt},
 m9/.style={mk,diamond,minimum size=8.5pt},
 m8/.style={mk,regular polygon,regular polygon sides=3,minimum size=8pt},
 m10/.style={mk,star,star points=5,star point ratio=2.3,minimum size=9pt},
 m6/.style={mk,regular polygon,regular polygon sides=5,minimum size=7.5pt},
 m7/.style={mk,rectangle,minimum size=5.5pt},
 cell/.style={draw=gray!65},
 excluded/.style={pattern=north east lines,pattern color=gray!60}]
 % layers: U above, D below, the middle band split by its diagonal into two halves
 \fill[green!12] (.5,10.5) rectangle (15.5,15.5);
 \fill[orange!12] (.5,.5) rectangle (15.5,5.5);
 \fill[blue!14] (.5,5.5) -- (.5,10.5) -- (15.5,10.5) -- cycle;
 \fill[blue!5] (.5,5.5) -- (15.5,5.5) -- (15.5,10.5) -- cycle;
 \draw[gray!65] (.5,.5) rectangle (15.5,15.5);
 \draw[gray!65] (.5,5.5)--(15.5,5.5) (.5,10.5)--(15.5,10.5);
 % the threshold tau = floor(c/2): the diagonal of the middle band
 \draw[threshold] (.5,5.5) -- (15.5,10.5) node[right,font=\small] {$\tau$};
 % points: dots for U and D, one shape per middle point
 \foreach \x/\y in {1/12,3/15,5/11,10/13,14/14,7/5,8/2,12/4,13/1,15/3}{
  \fill (\x,\y) circle[radius=2pt];
 }
 \node[m9] at (2,9) {}; \node[m8] at (4,8) {}; \node[m10] at (6,10) {};
 \node[m6] at (9,6) {}; \node[m7] at (11,7) {};
 \node[right] at (15.9,13) {$U$};\node[right] at (15.9,8) {$M$};\node[right] at (15.9,3) {$D$};
 % axes
 \draw[->,gray] (-.9,-.9)--(16.6,-.9) node[anchor=north east,yshift=-9pt,font=\small]{position};
 \draw[->,gray] (-.9,-.9)--(-.9,16.5) node[above]{value};
 \foreach \v in {1,...,15}{
  \draw[gray] (\v,-1.05)--(\v,-.75);
  \draw[gray] (-1.05,\v)--(-.75,\v);
 }
 \foreach \v in {1,3,...,15}{
  \node[below,font=\footnotesize] at (\v,-1.0) {$\v$};
  \node[left,font=\footnotesize] at (-1.0,\v) {$\v$};
 }
 % legend: the five middle points by value, then the other points
 \node[m9] at (1.0,-4.0) {}; \node[right,font=\footnotesize] at (1.4,-4.0) {$9$};
 \node[m8] at (3.2,-4.0) {}; \node[right,font=\footnotesize] at (3.6,-4.0) {$8$};
 \node[m10] at (5.4,-4.0) {}; \node[right,font=\footnotesize] at (5.8,-4.0) {$10$};
 \node[m6] at (8.2,-4.0) {}; \node[right,font=\footnotesize] at (8.6,-4.0) {$6$};
 \node[m7] at (10.3,-4.0) {}; \node[right,font=\footnotesize] at (10.7,-4.0) {$7$};
 \fill (12.5,-4.0) circle[radius=2pt]; \node[right,font=\footnotesize] at (12.9,-4.0) {other};
 % rank-cell grids of the two maps: rows 1..5 of the decreasing rank, the threshold
 % as a horizontal line, cells above it and cells below the double-ended arrows excluded
 \begin{scope}[xshift=6.1cm,yshift=.1cm,x=.85cm,y=.62cm]
  \node[anchor=south,font=\small] at (1.5,4.8) {first map $(\estrand,\de{P})$};
  \fill[excluded] (.5,.5) rectangle (1.5,1.5);
  \fill[excluded] (1.5,.5) rectangle (2.5,2.5);
  \fill[excluded] (.5,3.5) rectangle (2.5,4.5);
  \fill[blue!14] (.5,1.5) rectangle (1.5,3.5);
  \fill[blue!14] (1.5,2.5) rectangle (2.5,3.5);
  \draw[cell] (.5,.5) rectangle (2.5,4.5);
  \draw[cell] (1.5,.5)--(1.5,4.5);
  \foreach \r in {1.5,2.5,3.5}{\draw[cell] (.5,\r)--(2.5,\r);}
  \draw[threshold] (.3,3.5) -- (2.8,3.5) node[right,font=\small] {$\tau$};
  \draw[->,gray] (0,0)--(3,0) node[right,font=\small] {$\estrand$};
  \draw[->,gray] (0,0)--(0,4.9) node[left,font=\small] {$\de{P}$};
  \foreach \z in {1,2}{\node[below,font=\footnotesize] at (\z,0) {$\z$};}
  \foreach \z in {1,...,4}{\node[left,font=\footnotesize] at (0,\z) {$\z$};}
  \draw[dim] (.6,.5)--(.6,1.5); \node[font=\footnotesize,inner sep=1pt,fill=white] at (1.08,1) {$|\Ustar{1}|$};
  \draw[dim] (1.6,.5)--(1.6,2.5); \node[font=\footnotesize,inner sep=1pt,fill=white] at (2.08,1.5) {$|\Ustar{2}|$};
  \node[m9] at (1,2) {};\node[m8] at (1,3) {};\node[m10] at (2,3) {};
 \end{scope}
 \begin{scope}[xshift=9.45cm,yshift=.1cm,x=.85cm,y=.62cm]
  \node[anchor=south,font=\small] at (1.5,4.8) {second map $(\sstrand,\ds{P})$};
  \fill[excluded] (.5,.5) rectangle (2.5,2.5);
  \fill[excluded] (.5,3.5) rectangle (2.5,4.5);
  \fill[blue!5] (.5,2.5) rectangle (2.5,3.5);
  \draw[cell] (.5,.5) rectangle (2.5,4.5);
  \draw[cell] (1.5,.5)--(1.5,4.5);
  \foreach \r in {1.5,2.5,3.5}{\draw[cell] (.5,\r)--(2.5,\r);}
  \draw[threshold] (.3,3.5) -- (2.8,3.5) node[right,font=\small] {$c-\tau$};
  \draw[->,gray] (0,0)--(3,0) node[right,font=\small] {$\sstrand$};
  \draw[->,gray] (0,0)--(0,4.9) node[left,font=\small] {$\ds{P}$};
  \foreach \z in {1,2}{\node[below,font=\footnotesize] at (\z,0) {$\z$};}
  \foreach \z in {1,...,4}{\node[left,font=\footnotesize] at (0,\z) {$\z$};}
  \draw[dim] (.6,.5)--(.6,2.5); \node[font=\footnotesize,inner sep=1pt,fill=white] at (1.08,1.5) {$|\Dstar{1}|$};
  \draw[dim] (1.6,.5)--(1.6,2.5); \node[font=\footnotesize,inner sep=1pt,fill=white] at (2.08,1.5) {$|\Dstar{2}|$};
  \node[m7] at (1,3) {};\node[m6] at (2,3) {};
 \end{scope}
\end{tikzpicture}
\caption{Example~\ref{ex:split-shapes} with $\tau=\floor{c/2}=3$.
Lemma~\ref{lem:triangle} bounds the points in the middle layer
together with points in the upper layer
by mapping points in $M$ to boxes on the right.
For the points on the upper left of the line marked by $\tau$,
there are 6 boxes
in the middle graph that are under
the threshold~$\tau$,
and two points remaining in $U$ to be counted separately.
For the points on the lower right of the line,
we count 6 boxes from the right graph
and two more points in the lower layer
that correspond to one decreasing strand in $D$.
Overall we count 6+2+6+2=16 points,
which is greater than 15, the actual number of points.}
\label{fig:split-shapes}
\end{figure}
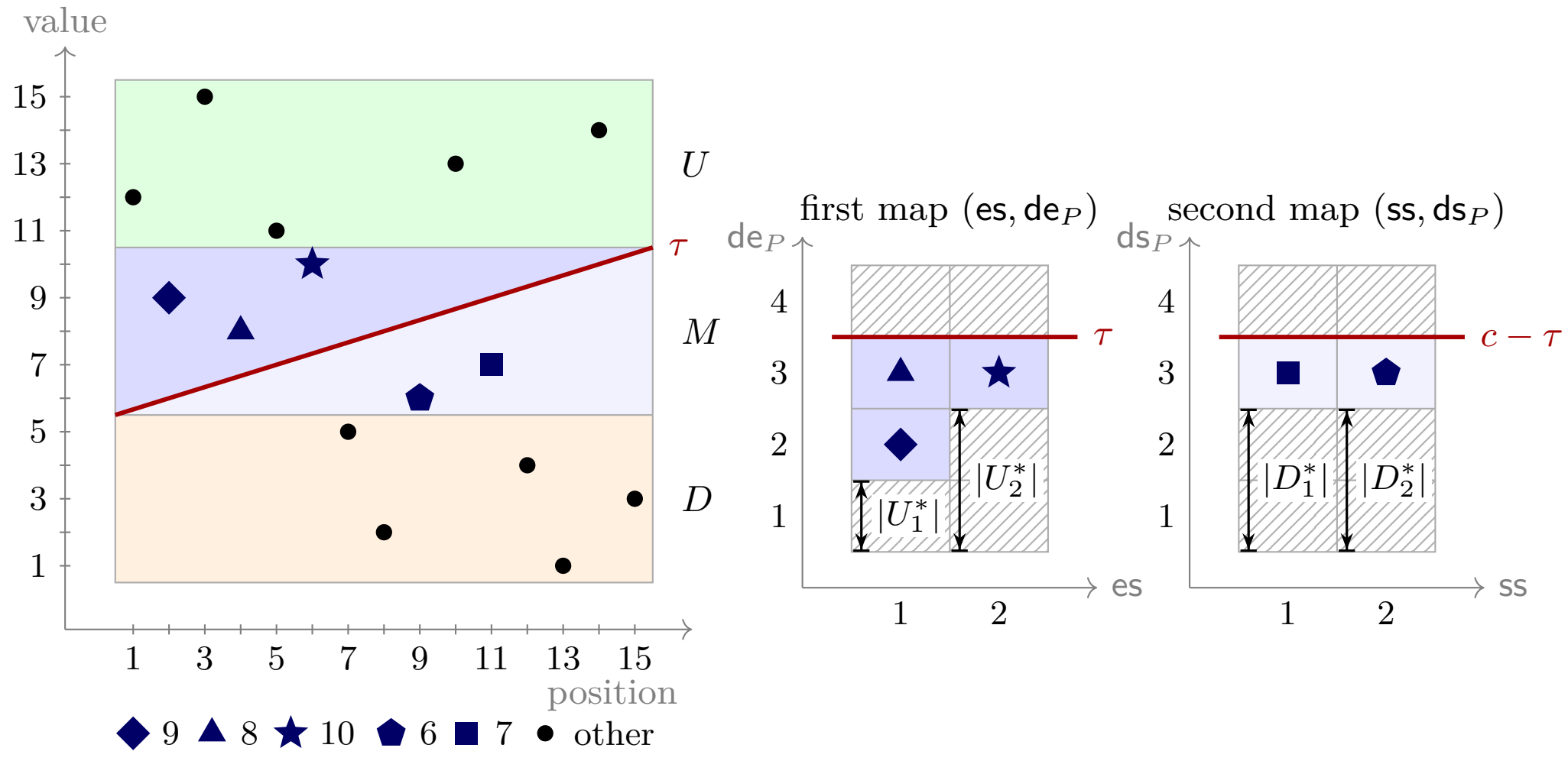

\begin{example}\label{ex:split-shapes}
Let $c=6$, $w=4$, and $P=(12,9,15,8,11,10,5,2,6,13,7,4,1,14,3)$,
with the upper cut between values 10 and 11,
and the lower cut between values 5 and 6.
We visualise such a sequence
and the following argument
in Figure~\ref{fig:split-shapes}.
Then, we have $U=(12,15,11,13,14)$, $M=(9,8,10,6,7)$, and~$D=(5,2,4,1,3)$. The
tightened labels of $M$ are
$(\estrand,\sstrand)=(1,2)$, $(1,2)$, $(2,1)$, $(1,2)$, $(2,1)$ in
positional order; assign $(1,2)$, $(2,2)$, $(2,1)$, $(3,1)$, $(4,1)$ to
the points of $U$ and $(1,3)$, $(2,3)$, $(2,1)$, $(3,2)$, $(3,1)$ to the
points of~$D$, in positional order. With these labels,
$P$ admits a \layerrep{6}{4} with
$\LIS(P)=4$, $\LDS(P)=6$, and $h=\LIS(M)=2$.
Note that we have $c+h=8=2w$.

Recall the partition of $U$ and $D$
into decreasing chains~$U_i$ and $D_j$
by the equivalence of ending and starting strands,
respectively.
Then, we have $U_1=\{12\}$,
$U_2=\{15,11\}$, $U_3=\{13\}$, and $U_4=\{14\}$
based on the ending strands,
and $D_1=\{4,3\}$, $D_2=\{1\}$, $D_3=\{5,2\}$, and $D_4=\varnothing$
based on the starting strands.
Thus, we satisfy $|U_i|\le w+1-i$ and $|D_j|\le w+1-j$
for each ending and starting strand~$i$ and $j$.
Furthermore,
we have $|\Ustar{1}|=1$, $|\Ustar{2}|=2$, and $|\Dstar{1}|=|\Dstar{2}|=2$.

The remaining subsequence with values~$9,8,10,6,7$ equals the middle layer.
The middle points have the following decreasing ranks in $P$:
\begin{center}
\begin{tabular}{c|*{5}{wc{1.6em}}}
 position of point~$p$&2&4&6&9&11\\
 value of point~$p$&9&8&10&6&7\\\hline
 $\de{P}{p}$&2&3&3&4&4\\
 $\ds{P}{p}$&5&4&4&3&3\rlap{.}
\end{tabular}
\end{center}
Choose $\tau=\floor{c/2}=3$,
which is exactly the choice of $\tau$ in Proposition~\ref{prop:joint}.
The middle points $9,8,10$ have $\de{P}\le\tau$,
and we count by mapping them to $(\estrand,\de{P})$,
namely $(1,2)$, $(1,3)$, and $(2,3)$.
Note that $|\Ustar{\estrand}|<\de{P}\le\tau$.
The points $6,7$ have $\de{P}>\tau$, hence
$\ds{P}\le c-\tau=3$ by~\eqref{eq:rank-sums}.
We count these points
by mapping them to $(\sstrand,\ds{P})$, namely $(2,3)$ and $(1,3)$.
Again, note that $|\Dstar{\sstrand}|<\ds{P}\le c-\tau$.
The four summands of~\eqref{eq:split} are $5$, $(3-1)+(3-2)=3$, $5$,
and $(3-2)+(3-2)=2$, so the bound is $15=|P|$.
Lemma~\ref{lem:triangle} bounds each half of the count by
$w(w+1)/2=10$, and Proposition~\ref{prop:joint} gives $|P|\le20$.

\end{example}

\FloatBarrier
\section{Deriving \texorpdfstring{$\extremalf{k}{n}$}{f\_k(n)} from \texorpdfstring{$\extremalN{k}{L}$}{N\_k(L)}: proof of Theorem~\ref{thm:main}}\label{app:inversion}
We prove Theorem~\ref{thm:main} from Theorem~\ref{thm:size} and
Lemma~\ref{lem:duality} by inverting the size function, keeping track
of the integer boundaries. For every $n$, we also provide
an $n$-point permutation whose longest rollercoaster has exactly
$\extremalf{k}{n}$ points.
\begin{proof}[Proof of Theorem~\ref{thm:main}]
For $n\le q^2$, any $n$ points of $\gridperm{q}{q}$ give an input with no
rollercoaster, so $\extremalf{k}{n}=0$.

Suppose $q^2<n\le2q^2$, and let
$t=\floor{\sqrt{2q^2-n}}$ and $\ell=2q-t$.
Then, we have $0\le t<q$ and
$2q^2-(t+1)^2<n\le2q^2-t^2$.
The formula for $N_k$, including its endpoints, therefore gives
$\extremalN{k}{\ell-1}=2q^2-(t+1)^2<n
 \le2q^2-t^2=\extremalN{k}{\ell}$.
By Lemma~\ref{lem:duality}, we have $\extremalf{k}{n}=\ell$.
For the sequence achieving $\extremalf{k}{n}=\ell$,
recall that $\sevenperm{q}{t}$
has a longest $\ell$-point rollercoaster.
We choose any $n$-point subsequence of $\sevenperm{q}{t}$
that contains this $\ell$-point rollercoaster
and delete all other points.
This gives a sequence of $n$ points
whose longest rollercoaster has $\ell$ points.

Finally, for $n>2q^2$, let $m=\ceil{n/q}$. Since $m-1\ge2q$,
we have $\extremalN{k}{m-1}=q(m-1)<n\le qm=\extremalN{k}{m}$.
Again Lemma~\ref{lem:duality} gives $\extremalf{k}{n}=m$.
For the sequence achieving $\extremalf{k}{n}=m$,
we use a similar argument to the
$q^2<n\le2q^2$ case.
Consider $\gridperm{q}{m}$; the longest rollercoaster
of $\gridperm{q}{m}$ has length $m$.
Choose a subsequence of length~$n$ of $\gridperm{q}{m}$
that contains this rollercoaster
and delete everything else.
This yields a sequence of $n$ points
whose longest rollercoaster has $m$ points.
\end{proof}

\section{Computing a rollercoaster meeting the guarantee}\label{app:algorithm}
We explain the algorithm that takes an $n$-point
sequence~$S=(x_1,\ldots,x_n)$ of distinct elements
and an integer~$k\ge 3$ as input
and computes a rollercoaster subsequence
with length at least $\extremalf{k}{n}$.
We also prove its correctness and running time,
which is stated in Theorem~\ref{thm:algorithm}.
Specifically, the running time of the algorithm for $n>q^2$ is
$O(n\log\extremalf{k}{n})$, and we use $O(n)$ space,
which includes the links used for backtracking the solutions.
We work in the standard word-RAM model with word size
$\Omega(\log n)$ and allow constant-time comparisons between input elements.
Although we conceptually treat the input as a permutation,
we do not replace the elements with their ranks in the input
and compare the elements directly.

If $n\le q^2$, we simply output the empty sequence
since $\extremalf{k}{n}=0$.
Note that there may exist a rollercoaster in the input,
but we nevertheless output one that has $\extremalf{k}{n}=0$ elements.
For $q^2<n\le2q^2$,
we compute the starting and ending chain ranks
and either find a single chain of length $\extremalf{k}{n}$,
or join two chains, each of length at least $q+1$
at a single point.
One of the two types of rollercoaster is guaranteed to exist
by $\extremalf{k}{n}\le 2q$.
The idea is to bound the value of the chain rank with $\extremalf{k}{n}$;
then applying a modified version of the patience sorting algorithm
which we describe shortly after.
For $n>2q^2$, the algorithm first identifies points at which a
rollercoaster of at least a specified initial length ends.
This can also be done by applying the modified patience sorting algorithm
a constant number of times.
It then processes
these marked points in input order and assigns them to classes
satisfying two invariants.
Here, the invariants guarantee the existence
of rollercoasters ending at a point in class $i$
with length proportional to $i$.
Once we put a point in the largest class,
we backtrack the algorithm to retrieve and return a rollercoaster
with length $\extremalf{k}{n}$.

\subparagraph*{Patience sorting.}
We now describe the patience sorting algorithm~\cite{AldousD99}
for completeness.
Given a sequence of $n$ distinct elements,
we first initialise an array of $n$ empty stacks
and push the first input element to the leftmost stack.
Then, we repeat the following
starting from the second input element.
We find the leftmost nonempty stack whose top element
is no less than the current element,
and insert the current element there.
If no such stack exists,
we insert into the leftmost empty stack.
This maintains the property that
the top elements of each stack are in increasing order.
After we insert all elements into the stack,
a point is stored in the $i$th stack
if the longest increasing subsequence
ending there has length~$i$.
Therefore, the number of nonempty stacks is equal
to the length of the longest increasing subsequence
of the input sequence.
From the invariant, we can search for the stacks
using binary search on the top elements of the stacks,
and therefore this gives us an algorithm
for computing the length of the longest increasing subsequence
ending at each point in $O(n\log n)$ time.
Finally, we can return a longest subsequence
ending at some given point
by maintaining references for each element at the time of sorting;
whenever some element is inserted in a stack,
we remember which element was stored at the top
of the stack directly to the left.
Then, we can retrieve one of the longest increasing subsequences
ending at some point
by backtracking the references, starting at the given point.

\subsection{Finding shorter rollercoasters}\label{app:scans}
The underlying recurrence in the patience sorting algorithm
is that the length of the longest increasing subsequence ending at some position
is one plus the length of the longest increasing subsequence
ending at all previous positions with elements smaller than the current one.
We use the same extension rule, with
two differences. First, a subsequence may start only at specified positions,
where each position is also provided an integer
representing some initial progress up to that position.
For example, each point may be given
the length of a longest rollercoaster ending there
that consists up to some constant number of runs.
Also, the computation does not distinguish lengths once they reach a
specified cap. We call one execution of the modified
patience sorting algorithm
a \emph{scan}.

We now describe a scan in detail.
Let~$V$ be a set of starting positions,  $C$ be an integer cap with
$1\le C\le n$, and $\startlen{v}\in\{1,\ldots,C\}$
be the initial progress for each $v\in V$.
For each position $p$, let
$\scanlen{p}$ be the maximum of
the initial progress of $v$
plus the longest increasing subsequence
that starts at $v$ and ends at $p$ minus one, or $-\infty$ if none
exists,
where the maximum is taken over points~$v\in V$.
Note that the minus one term exists because $v$ is counted twice,
once in the initial progress and once in the longest increasing subsequence.
Then, the scan computes $\min\{\scanlen{p},C\}$
for each position~$p$,
which we call the \emph{capped progress at $p$}.
A scan must also allow backtracking
for the retrieval of some position~$v$
and some subsequence~$X$ starting at $v$ and ending at $p$
that achieves $\min\{\scanlen{p},C\}\le \startlen{v}+|X|-1$.
We allow this by maintaining predecessor links,
where the link for $p$ remembers
one of the points that contributed in computing
the capped progress at $p$.

\begin{lemma}\label{lem:prescribed-starts}
The capped progress and the predecessor links
for all positions can be computed in
$O(n\log(C+1))$ total time and $O(n)$ space. The same bounds hold for
decreasing subsequences and for scans from right to left.
\end{lemma}
\begin{proof}
We process the positions from left to right.
At position~$p$, we take the maximum over
one plus the maximum of the capped progress at positions~$v<p$
with $x_v<x_p$,
and also $\startlen{p}$ if $p\in V$.
If no such $v$ exists and $p$ is not in $V$,
then we set the capped progress at $p$ to $-\infty$.
Otherwise, we take the minimum between the computed value
and $C$, and assign this value as the capped progress at $p$.
We also remember one of the positions that contributed
to computing the maximum value, and assign that to the predecessor link
for $p$.
If $\startlen{p}$ is the maximum,
we simply remember $p$, which represents that
the subsequence begins at $p$.

For the correctness of the recurrence,
every increasing subsequence ending at $p$ either starts at $p$ or
extends another increasing subsequence
that ends at an earlier point with smaller value. Moreover,
capping after every addition of one is equivalent to capping only the
final maximum value. Therefore, we correctly compute
$\min\{\scanlen{p},C\}$ for each point~$p$.
Furthermore,
if $p$ is recorded in the predecessor link,
the initial progress~$\startlen{p}$ contributes to
computing the capped progress at $p$,
which means that the increasing subsequence consists of a single point~$p$.
Otherwise, whenever we record another position~$p'<p$,
then $\min\{\scanlen{p'},C\}+1$ is at least $\min\{\scanlen{p},C\}$.
Recursively applying this argument,
we will eventually reach some position~$v$ that records itself,
which proves that $\min\{\scanlen{p},C\}\le \startlen{v}+|X|-1$,
where $X$ is the sequence of positions that are backtracked.

Since we allow arbitrary initial progresses,
we can no longer maintain the predecessors in a range of contiguous stacks.
Therefore, we achieve logarithmic search and insertion
with a balanced binary search tree
that holds items that correspond to the topmost elements of nonempty stacks
in the patience sorting algorithm.
Specifically, the items are pairs of positions
and the capped progress at that position.
They are sorted in increasing order
with respect to the capped progress.
Then, whenever we try to insert a pair~$(p,\min\{\scanlen{p},C\})$
to the search tree,
we find the item in the search tree
with the largest input element below $x_p$.
If the progress in this item is at least the progress
of the new item,
we discard the item we attempted to insert.
Otherwise,
we remove all pairs~$(v,\min\{\scanlen{v},C\})$ from the search tree
whenever $x_v>x_p$
and the capped progress at $v$ is no larger than
that at $p$.
Such a policy gives the search tree two invariants:
first, the capped progress of each item is unique,
where the position associated with the progress
has the smallest possible input element,
and second, the positions are stored in increasing order
with respect to the corresponding elements.

We can now describe the details of the algorithm
at each position of the input.
At each position,
we first find the item in the search tree
with the largest input element below $x_p$.
If no such item exists, then
no position~$v<p$ satisfies $x_v<x_p$.
In this case, we try to insert $(p,\min(\startlen{p},C))$
if $p\in V$, or do not insert otherwise.
On the other hand,
let $(v,g)$ be the pair
where $x_v$ is the largest element below $x_p$
among the positions in the search tree.
If $p\in V$, we try to insert $(p,\min(\max(\startlen{p},g+1),C))$,
and otherwise try to insert $(p,\min(g+1,C))$.
Furthermore, if $\min(g+1,C)$ or $\startlen{p}$
is the capped progress at $p$,
we respectively remember $v$ or $p$ with the predecessor link.

Since each item in the search tree
has a unique capped progress, we have at most $C+1$ items
in the search tree.
Thus, each search, insertion, or deletion takes
$O(\log(C+1))$ time. Every processed position incurs at most one
insertion and at most one eventual deletion, so the total time is
$O(n\log(C+1))$. Furthermore,
the predecessor links occupy $O(n)$ space.
Note that we can reverse the comparisons or the
processing order
to compute the capped progress
for decreasing sequences
or sequences that start at each point.
\end{proof}

Suppose first that $q^2<n\le2q^2$. Our goal is to find
a rollercoaster with at least
$\ell=\extremalf{k}{n} =2q-\floor{\sqrt{2q^2-n}}\in[q+1,2q]$ points.
We compute the
increasing and decreasing lengths ending and starting at each point,
each value capped at $2q$,
with four scans using Lemma~\ref{lem:prescribed-starts}.
These scans have every position in the starting set,
and for each position the initial progress is set to one.
For example, we use a scan from left to right
for increasing sequences to compute the lengths of
increasing chains ending at each point.
Note that the cap is at most $n$ because $2q\le q^2<n$.

If a capped progress is at least $\ell$ at some position,
we retrieve and return the corresponding chain
by backtracking the predecessor links. Otherwise,
Theorem~\ref{thm:main} guarantees a rollercoaster of at least $\ell$
points with at least two runs. At one of its turning points, a chain
of at least $q+1$ points ends and another such chain starts.
We can already identify such a point with these two properties
using the four scans. Select $q+1$ points
from each corresponding chain while retaining their common endpoint.
Then, their
concatenation has $2q+1$ points.
This sequence is a rollercoaster because the two
directions either agree, giving one run, or differ, giving two runs
of length $q+1$. This procedure takes $O(n\log q)$ time and $O(n)$
space, including the time it takes to retrieve the output.

Assume from now on that $n>2q^2$.
Now our goal is to find a rollercoaster
with at least $L=\ceil{n/q}>2q$ points.
Then, there is a unique integer~$\beta\in\{2q+1,\ldots,3q\}$
that is congruent to $L$ modulo $q$.
Let $T=(L-\beta)/q$,
and then we have
$L=\beta+Tq$.
In the first half of the algorithm,
we mark precisely those positions $p$ at which a rollercoaster
of at least $\beta$ points ends.
For each marked point, the computation
also records how to retrieve one such rollercoaster.
Then, once the marking is complete,
we extend the rollercoasters ending at the marked points
by $q$ points each step
and return a rollercoaster whenever we extend
some rollercoaster $T$ times.

\begin{lemma}\label{lem:seeds}\label{app:seeds}
We can mark all positions~$p$ with $\rcend{p}\ge \beta$
and compute predecessor links allowing the retrieval of
such rollercoasters
in $O(n\log q)$ time and $O(n)$ space.
\end{lemma}
\begin{proof}
The algorithm consists of three phases. 
In the first phase,
we identify the ending ranks capped at $\beta$
for all positions in the input.
Then, in the second phase,
we find the positions where
a rollercoaster with at least two runs can end.
Finally, in the third phase,
we find the positions where a rollercoaster
with at least $\beta$ points can end.

\subparagraph*{Phase One.}
We compute the ending ranks capped at $\beta$ for each position.
We use two scans, each with cap $\beta$ and starting length 1
for every position in the input.
Since $\beta\le 3q$, this can be done in $O(n\log q)$ time
and $O(n)$ space
by Lemma~\ref{lem:prescribed-starts}.

\subparagraph*{Phase Two.}
We compute the positions where a rollercoaster with at least two runs can end.
Consider the last two runs in such a rollercoaster.
Both of the last two runs have at least $q+1$ points.
We find such runs using two scans, one for each direction of the final run.
For the case when the final run is increasing or decreasing,
we let the starting set be the positions
whose decreasing and increasing ending chain ranks are at least $q+1$,
respectively.
In both cases, the initial progress is set to one
and the cap is set to $\beta$.
Thus, we return the positions with progress at least $q+1$ in either scan.
This completes the computation
of the positions where a rollercoaster with two or more runs can end.
Then, Phase Two can also be done in $O(n\log q)$ time
and $O(n)$ space
by Lemma~\ref{lem:prescribed-starts}.
Note that we can also remember the direction of the final run
for each returned position; increasing, decreasing or both.

\subparagraph*{Phase Three.}
We compute the positions where a rollercoaster
of length at least $\beta$ can end.
For rollercoasters consisting of a single run,
we access the ending ranks computed in Phase One
and mark the positions whose capped progress equals $\beta$.
For rollercoasters with two runs,
we find the endpoints using two scans,
one for each direction of the final run.
For the case when the final run is increasing,
we let the starting set be the positions
whose decreasing ending chain ranks are at least $q+1$
and let the cap be $\beta$.
This time, however, we let the starting progress
for each position~$p$
be the minimum between its decreasing ending chain rank
and $\beta-q$;
the intuition for $\beta-q$ is that
the increasing sequence must have at least $q+1$ points
to be a part of a rollercoaster.
In other words, for decreasing sequences
with more than $\beta-q$ points,
we need $q$ more increasing points
in order to form a rollercoaster,
and thus we can only count a progress of $\beta-q$
towards completing a rollercoaster of length at least $\beta$.
The scan for finding endpoints for rollercoasters with two runs
whose final run is decreasing can be done analogously,
and we mark the positions
whose capped progress equals $\beta$
in either scan.

For rollercoasters with three or more runs,
consider the last three runs in such a rollercoaster.
Since $\beta\le 3q$, any three consecutive runs constitute a rollercoaster
with at least $\beta$ points.
Therefore, we use two additional scans for the direction of the last run
in such a rollercoaster.
For finding increasing and decreasing final runs,
we let the starting positions be the set of positions
found in Phase Two
using decreasing and increasing final runs, respectively.
For these positions, assign 1 as the initial progress
and let $q+1$ be the cap.
Then, the positions with progress $q+1$ at the end of each scan
have a rollercoaster with at least three runs ending there.
We mark these positions as well.
Since Phase Three uses four scans with cap $O(q)$
on at most $n$ points,
we can compute all such positions in $O(n\log q)$ time
and $O(n)$ space
by Lemma~\ref{lem:prescribed-starts}.

\subparagraph*{Summing up.} The overall running time for computing
the endpoints of rollercoasters with at least $\beta$ points
is $O(n\log q)$,
and the overall procedure uses $O(n)$ auxiliary space
for storing the predecessor links for each scan.
Furthermore, note that each scan finds a run in the opposite direction
of the previous run.
Therefore, at any position marked in Phase Three,
we can follow the predecessor links associated with the scan
that marked that position
and arrive at some other position referring to itself.
If the corresponding scan uses points resulting from Phase One or Phase Two,
we repeat the same backtracking method
in each corresponding scan.
Otherwise, we retrieve the sequence of positions
that are visited while backtracking
in increasing positional order.
Note that a single backtracking can be done in $O(n)$ time,
since there are only 8 scans
and we visit a position at most two times
over the entire course of the backtracking.
\end{proof}

\subsection{Extending the stored rollercoasters}\label{app:classes}
Let us now assume that we marked all end positions
for rollercoasters with at least $\beta$ points.
The next step is to assign each marked position
to disjoint classes $\class{0},\class{1},\ldots,\class{T}$. 
The points in each class are stored
in input order, so it is a subsequence of $S$.
Let us write $\stored{p}$ for the rollercoaster
we can retrieve by backtracking at a point $p$
using
Lemma~\ref{lem:seeds}. We maintain two
invariants:
\begin{enumerate}
 \item No class contains a monotone subsequence of $q+1$ points.
 \item If $p\in\class{h}$, then $\stored{p}$ has at least
       $\beta+hq$ points.
\end{enumerate}
We emphasise that a class is only a set of marked positions; it is not itself a
rollercoaster. The first property gives $|\class{h}|\le q^2$ by
Lemma~\ref{lem:ranks}, and the second gives a length guarantee for each
rollercoaster ending at positions in that class.
Once a position is placed in $\class{T}$,
the second invariant guarantees
a rollercoaster with $L=\beta+Tq$ points
ending at the position.

We say that a class \emph{accepts} a marked point $p$
if no $(q+1)$-point chain exists in the class
after adding $p$ to the class.
Otherwise, we say that the class \emph{rejects} $p$.
Note that an empty class accepts every point.
Before visiting any marked position~$p$,
we initialise the sets~$\class{0},\ldots,\class{T}$
as empty.
Then, for each marked position~$p$
in increasing positional order,
our goal is to find an integer~$h$
such that $\class{h}$ accepts $p$ and $\class{h-1}$ rejects $p$.
If no such $h$ exists,
then we let $h=0$. The next subsection explains how to
efficiently find such an $h$.

Assume that there exists such a positive $h$.
Then, the class~$\class{h-1}$ contains no monotone subsequence of
$q+1$ points by the first invariant.
Hence any such subsequence created by inserting $p$
must use $p$ as its last point.
Choose such a sequence~$a_0,a_1,\ldots,a_{q-1},p$
with $p$ points,
where $a_0,\ldots,a_{q-1}\in\class{h-1}$ and the values
$x_{a_0},\ldots,x_{a_{q-1}},x_p$ are monotone.
By the second invariant,
the rollercoaster~$\stored{a_0}$ has at least
$\beta+(h-1)q$ points.
Then, we add exactly $q$ later points
to $\stored{a_0}$ when we extend it using $a_1,\ldots,a_{q-1},p$.
If the direction of the extended sequence agrees with its
last run, that run becomes longer. Otherwise the new run starts at
$a_0$ and has $q+1$ points. This gives $\stored{p}$ and proves that
the second invariant holds. Furthermore,
inserting $p$ into the accepting class~$\class{h}$ proves that
the first invariant holds. Both properties therefore hold by induction over the
marked points.
When $h=T$, retrieve and return $\stored{p}$ after inserting $p$.
Otherwise insert $p$ into $\class{h}$ and continue.

\subparagraph*{Why we eventually insert a marked position into $\class{T}$.}
Let us count the number of unmarked positions.
The subsequence of unmarked positions contains no
rollercoaster of length at least $\beta$: otherwise its right endpoint would also be
a marked point. The extremal formula therefore bounds the
number of these points by
$\extremalN{k}{\beta-1}=q(\beta-1)$.
Since $L=\ceil{n/q}$, we have $n>q(L-1)$, and hence
\[
 \#\{\text{marked points}\}
 \ge n-q(\beta-1)
 >q(L-1)-q(\beta-1)=Tq^2.
\]
The first $T$ classes cannot contain all these points,
because each class has at most $q^2$ points
from the first invariant.
Some marked point~$p$ must therefore be put in $\class{T}$, which guarantees that
$\stored{p}$ has at least $L$ points. For $T=0$, the same
inequality guarantees a marked point in $\class{0}$.
In this case, the algorithm backtracks on the
first point and returns the retrieved subsequence.

For example, $q=3$ and $n=40$ give $L=14$, $\beta=8$, and $T=2$.
At least $40-3(8-1)=19$ points are marked, whereas two classes
can hold at most $18$ points. The first point assigned index $2$ thus
has a corresponding rollercoaster of at least $14$ points.

\subsection{Maintaining and choosing the classes}\label{app:class-implementation}
We now describe how we can decide acceptance of a marked position
in a given class. Specifically, we decide
whether or not the given point creates a chain of $q+1$ points
in a class.

\begin{lemma}\label{lem:class-tails}
We can test whether or not a class accepts a position
in constant time. On rejection, we can set a predecessor link
that points to a monotone $(q+1)$-point subsequence in the class,
including the new position,
in constant time. On acceptance, we can add the given position
to that class
in $O(\log q)$ time. The data structures for all classes
and the links for backtracking use $O(m)$ space after $m$ total insertions.
\end{lemma}
\begin{proof}
Within each class, we maintain two instances
of the patience sorting algorithm
with $q$ stacks each,
one for increasing subsequences
and one for decreasing subsequences.
These instances allow backtracking
to retrieve the longest monotonic subsequence
within positions of the class
that ends at each position.
Thus, we can check whether or not
the new position~$p$ creates a chain of length $q+1$
by simply checking the top element of the $q$th stack
in both of the instances;
no chain exists after insertion
if,
for the instance corresponding to increasing and decreasing subsequences,
the $q$th stack is empty
or the position~$v$ on the top of the $q$th stack
satisfies $x_v > x_p$ and $x_v < x_p$,
respectively.
This can be tested in constant time.

On rejection,
we can remember the position in the $q$th stack in each class
depending on the direction of the chain.
This can be done in constant time.
On the other hand,
on acceptance,
we can find the leftmost stack
whose element at the position at the top of the stack
is greater than the element at the given position
in logarithmic time.
This can be implemented using binary search
on the top elements.
This completes the proof.
\end{proof}

Suppose we currently have $\mu$ nonempty classes. Treat the fictitious
$\class{-1}$ as rejecting and the next empty class $\class{\mu}$ as
accepting. Initially set $\mathrm{left}=-1$ and
$\mathrm{right}=\mu$. While these indices are not consecutive, test
$\class{j}$ for $j=\floor{(\mathrm{left}+\mathrm{right})/2}$.
An accepting answer replaces the right endpoint by $j$; a rejecting
answer replaces the left endpoint. Thus the left index always names a rejecting class and the right
index always names an accepting class. The number of class indices
strictly between them at least halves at every test. When they
become consecutive, $h=\mathrm{right}$ gives us one of
the desired values~$h$.
We emphasise that
this is a search for any adjacent rejecting and accepting classes,
not necessarily the first accepting class. The proof uses only the
acceptance of $p$ at $\class{h}$ and rejection at $\class{h-1}$,
so acceptance need not be
monotone in the class index. We perform
at most $O(\log(\mu+1))$ tests when $\mu\ge1$, and no tests when $\mu=0$,
including the constant-time setup.
Therefore, the search for an $h$ takes
$O(1+\log(\mu+1))$ time.

For example, let $q=2$ and
$S=(8,9,14,3,13,15,2,11,1,16,12,10,4,17,6,7,5)$.
Then, we have $L=9$, $\beta=5$, and $T=2$.
When the marked point of value $10$ is processed,
the class~$\class{0}=(2,11,1)$ accepts it,
the class~$\class{1}=(16,12)$ rejects it, and
the next empty class~$\class{2}$ accepts it. The rejection follows
from the decreasing subsequence $(16,12,10)$. The search nevertheless
chooses $\class{0}$ correctly. The next marked point has value $4$;
both existing classes reject it, so it reaches index $2$ and the
algorithm returns its corresponding rollercoaster.

The following pseudocode specifies the algorithm
for $n>2q^2$.
Here, $\stored{p}$ denotes a link from $p$
to the second-to-last point
in a $(q+1)$-point chain ending at $p$;
that is, $\stored{p}$ is the predecessor link
between marked points.

\FloatBarrier
\begin{figure}[htbp]
\centering
\begin{minipage}{\linewidth}
\small
\hrule\smallskip
\textbf{Long-range algorithm}, for $n>2q^2$
\begin{tabbing}
\quad\=\quad\=\quad\=\quad\=\kill
Compute $L=\ceil{n/q}$, $\beta\in[2q+1,3q]$ and $T\ge0$ with $L=\beta+Tq$.\\
Compute all marked points and their recorded rollercoasters; set $\mu\gets0$.\\
\textbf{for each} marked point $p$ in input order \textbf{do}\\
\> $\mathrm{left}\gets-1$; $\mathrm{right}\gets\mu$.\\
\> \textbf{while} $\mathrm{right}-\mathrm{left}>1$ \textbf{do}\\
\>\> $j\gets\floor{(\mathrm{left}+\mathrm{right})/2}$.\\
\>\> \textbf{if} $\class{j}$ accepts $p$ \textbf{then} $\mathrm{right}\gets j$; \textbf{else} $\mathrm{left}\gets j$.\\
\> $h\gets\mathrm{right}$.\\
\> \textbf{if} $h=0$ \textbf{then} $\stored{p}\gets$ the predecessor link for $p$ from Lemma~\ref{lem:seeds}.\\
\> \textbf{else}\\
\>\> $\stored{p}\gets$ the second to last point in a chain ending at $p$ within $\class{h-1}$.\\
\> \textbf{if} $h=T$ \textbf{then} backtrack from $\stored{p}$ and \textbf{return} the retrieved sequence.\\
\> \textbf{if} $h=\mu$ \textbf{then} update $\mu\gets\mu+1$.\\
\> Insert $p$ into $\class{h}$.
\end{tabbing}
\hrule
\end{minipage}
\end{figure}
\FloatBarrier

\subsection{Backtracking and complexity}
Starting at the point placed in $\class{T}$,
we alternate between using $q$ predecessor links
computed in the patience sorting instances
in the same direction
within each class,
and using a single $\stored{p}$ that gives us a position
in the previous class
until we reach the first point
of the $q$-point chain within $\class{0}$.
This procedure will visit
exactly $Tq+1$ points, including the point at which we started backtracking.
Now, we backtrack using the predecessor links
computed while marking the positions,
which will give us another $\beta-1$ points to visit.
Then, we output all the positions visited so far
in increasing positional order,
which is the reverse order we visited while backtracking.
Overall, this gives us a rollercoaster
of length at least $\beta+Tq=L$.
Note that in the backtracking phase,
we visit the positions in decreasing order
and we visit each position at most two times.
Thus, we can retrieve a rollercoaster of length at least $L$
in $O(n)$ time after we put a point in $\class{T}$.

The eight scans take $O(n\log q)$ time.
For each marked point, each acceptance test
is constant-time.
Therefore, we can complete the search for the integer~$h$
in $O(1+\log(T+1))$ time where the bottleneck is the binary search on $h$.
An insertion takes
$O(\log q)$ time by Lemma~\ref{lem:class-tails}. Since $q\ge2$ and
$q(T+1)=L-\beta+q<L$,
the work done per marked point is
$O(\log q+\log(T+1))=O(\log L)$.
Therefore, putting the marked points in each class takes overall
$O(n\log L)$ time. The data structures and predecessor links
use $O(n)$ space in total.

In the long range, $L=\extremalf{k}{n}$. In the short range,
$q<\ell=\extremalf{k}{n}$ and the four scans take $O(n\log q)$ time.
Including the $O(n)$ time for retrieving the rollercoaster,
the output gives $O(n\log\extremalf{k}{n})$ time and $O(n)$ space in both ranges.
This completes the proof of Theorem~\ref{thm:algorithm}.

\FloatBarrier

\section{Computing a longest rollercoaster}\label{opt:app:algorithm}
We give the algorithm and proof of Theorem~\ref{opt:thm:longest} in full.
The input is
$S=(x_1,\ldots,x_n)$, with $n\ge1$ distinct elements of a totally
ordered set, and an integer $k\ge3$. The output is the list of positions
of a longest $k$-rollercoaster, or a report that none exists.
If $k>n$, report nonexistence immediately.
So, from now on, we assume that $3\le k\le n$.

As before, we work in the standard word-RAM model with word size
$\Omega(\log n)$ and constant-time comparisons between input values.
All operations on input values are comparisons; no numerical encoding
of those values is required. Logarithms are binary.
The running time of the presented algorithm is $O(nk\log k)$ and the space usage is
$O(nk)$ (including the links needed to retrieve the subsequence).

The starting point is the $O(nk^2)$-time algorithm of Gawrychowski
et al.~\cite{GawrychowskiMS22}. We retain its main feature, the decomposition of the
input into contiguous segments, as well as its dynamic programming approach and recurrence.
The aforementioned decomposition limits the earlier dynamic programming states that need to
be considered when computing a new state. Our improvement is a faster search among these states, 
based on the combinatorial properties of the parts of this decomposition.
Appendix~\ref{opt:states} defines the decomposition and the
pseudocoaster lengths, then proves the recurrence. We prove
the required data-structure bound in Appendix~\ref{opt:frontier-section},
apply it in Appendix~\ref{opt:evaluation}, and account for the output
and all costs in Appendix~\ref{opt:complexity}.

\subsection{The decomposition and the dynamic program}\label{opt:states}
This subsection describes the decomposition and recurrence underlying
the first algorithm of Gawrychowski et al.~\cite{GawrychowskiMS22}.
The details of the faster data structure are independent of the
correctness of this recurrence.

\subparagraph*{Decomposing the input.}
Partition $S$ into contiguous, nonempty segments $P_1,\ldots,P_m$ as
follows. Starting at the beginning of the remaining suffix, an
odd-numbered segment is its shortest prefix containing an increasing
$k$-point subsequence. An even-numbered segment is its shortest prefix
containing a decreasing $k$-point subsequence. If the required
subsequence does not exist, take the remaining suffix as the last
segment. Write $\segidx{p}$ for the index of the segment containing
input position $p$.

We store the points of each segment in both input order and value order. The dynamic
program processes points in input order; the search structures are
initialised in value order.
The following patience-sorting computation~\cite{AldousD99} finds the
segments and partitions each into at most $k$ monotone subsequences.
During the processing of an odd-numbered segment, maintain the smallest last value of
an increasing subsequence of each length found so far. These last
values are the \emph{tails}; they increase strictly with the length.
For the next point, binary search finds the greatest tail smaller than
its value. The length of that tail's subsequence, plus one, is the
longest increasing subsequence length ending at the new point within
the segment; if no tail is smaller, this length is one. This length is
its increasing ending rank. Replace the tail for that rank by the new
value, or append a tail if the rank is new. End the segment when rank
$k$ is first reached.

Record the positions of each rank in a list, in input order. There
are at most $k$ nonempty lists, and each is decreasing. Indeed, if
positions $p<j$ in one list satisfied $x_p<x_j$, appending $x_j$
would extend an increasing subsequence ending at $p$ and force a
larger rank at $j$. For an even-numbered segment, use
decreasing ending ranks with the comparison order reversed. Its
lists are increasing. The final incomplete segment also has at most
$k$ lists. These lists partition each segment into monotone
subsequences. Each point requires $O(\log k)$ time, so computing this
decomposition and its monotone lists takes $O(n\log k)$ time and
$O(n)$ space. Retain each point's list index as well as the lists.

Reverse the decreasing lists within each segment and merge its at
most $k$ increasing lists with a heap. This sorts each segment by
value in $O(n\log k)$ total time and $O(n)$ space. The lists in
input order are retained separately; sorting does not change the
order in which the dynamic program processes the points.

Gawrychowski et al.~\cite{GawrychowskiMS22} prove the following
locality property for this decomposition. We call it the
\emph{locality lemma}.
\begin{lemma}\label{opt:locality}
If $p$ and $j$ are consecutive selected points in a run of a longest
$k$-rollercoaster and $j\in P_\ell$, then
$p\in P_{\ell-4}\cup\cdots\cup P_\ell$, with nonexistent segments
omitted.
\end{lemma}

Intuitively, in the rest of the algorithm, for each point~$p$, the algorithm computes the length of the longest subsequence ending at $p$
that is also a rollercoaster up to the second-to-last run; the last run in the subsequence may have less than $k$ points.
We call such a sequence a \emph{pseudocoaster}, and the details of this process are described in the next section.

\subparagraph*{What the pseudocoaster lengths mean.}
The intermediate subsequences satisfy the rollercoaster condition
except that their final run may have fewer than $k$ points. We use
two pseudocoaster length functions $\dptab{+}$ and $\dptab{-}$,
defined for positions $j\in[n]$ and final-run lengths $s=2,\ldots,k$.
The sign $\sigma\in\{+,-\}$ specifies an
increasing or decreasing final run, respectively. The pseudocoaster
length $\dptab[j,s]{\sigma}$ is the greatest length of a pseudocoaster ending
at position $j$ that satisfies these conditions:
\begin{itemize}
\item every run before the final run has at least $k$ points;
\item the final run has direction $\sigma$ and exactly $s$ points
when $s<k$, or at least $k$ points when $s=k$;
\item consecutive selected positions belong to segments whose
indices differ by at most four.
\end{itemize}
The length is $-\infty$ if no such pseudocoaster exists. We use
$\max\varnothing=-\infty$ and $1+(-\infty)=-\infty$.
Every finite length counts all selected points, not only the points
of the final run. In particular, a turning point is counted once.

The lengths $\dptab[j,k]{\sigma}$ describe rollercoasters. Every consecutive
pair of points of a rollercoaster belongs to one of its runs, so
Lemma~\ref{opt:locality} ensures that a longest rollercoaster obeys
the segment restriction imposed on the length functions. Consequently,
the largest finite length $\dptab[j,k]{\sigma}$, over both directions, is the
optimum. If all these lengths are $-\infty$, no $k$-rollercoaster exists.
This argument needs locality only for a longest rollercoaster;
the lengths for shorter final runs are explicitly defined with the
same segment restriction.

\subparagraph*{The lengths available before an extension.}
We compute the lengths position by position in increasing input order. Suppose that $p<j$ and
$0\le\segidx{j}-\segidx{p}\le4$. Appending $x_j$ to a subsequence
ending at $p$ extends an increasing final run only if $x_p<x_j$,
and a decreasing final run only if $x_p>x_j$.

For a final-run length $s$ and a direction $\sigma$, let
$\dppred{\sigma}{p}{s}$ be the greatest subsequence length available
at $p$ before appending a point in direction $\sigma$ to obtain
final-run length $s$. This is the length before the new point is appended,
whereas the pseudocoaster length computed by that update includes the new point.
The distinction determines what each search structure must store.
Write $\overline\sigma$ for the opposite direction.
The already computed lengths at $p$ give
\begin{equation}\label{opt:weights}
 \dppred{\sigma}{p}{s}=
 \begin{cases}
  \max\{1,\dptab[p,k]{\overline\sigma}\},&s=2,\\
  \dptab[p,s-1]{\sigma},&3\le s<k,\\
  \max\{\dptab[p,k-1]{\sigma},\dptab[p,k]{\sigma}\},&s=k.
 \end{cases}
\end{equation}
The first case starts from the singleton $p$, of length one, or
starts a new run after a run in the opposite direction of length
at least $k$. The middle case extends a final run of length $s-1$.
The last case extends a run of length $k-1$ or a run already of
length at least $k$. These cases are disjoint because $k\ge3$;
the middle case is empty when $k=3$.

For each final-run length $s$, take the largest length allowed by the position,
segment, and value comparisons, then add one for $j$:
\begin{equation}\label{opt:recurrence}
\begin{aligned}
 \dptab[j,s]{+}
   &=1+\max_{\substack{p<j,\ 0\le\segidx{j}-\segidx{p}\le4\\x_p<x_j}}
          \dppred{+}{p}{s},\\
 \dptab[j,s]{-}
   &=1+\max_{\substack{p<j,\ 0\le\segidx{j}-\segidx{p}\le4\\x_p>x_j}}
          \dppred{-}{p}{s}.
\end{aligned}
\end{equation}
At $j=1$ both maxima are empty, so every length at position~1 is
$-\infty$. The singleton at position~1 can nevertheless start a
two-point run later, through the first case of~\eqref{opt:weights}.

\begin{lemma}\label{opt:dp-correct}
Equations~\eqref{opt:weights} and~\eqref{opt:recurrence}, evaluated
in input order, compute the pseudocoaster lengths defined above. The
largest finite length $\dptab[j,k]{\sigma}$ is the longest $k$-rollercoaster
length; if all these lengths are $-\infty$, none exists.
\end{lemma}
\begin{proof}
Assume that all lengths at positions before $j$ have their stated meanings. Consider
any subsequence counted by $\dptab[j,s]{\sigma}$, and remove its
last point $j$. Let $p$ be the new endpoint.
When $s=2$, either $p$ is the only remaining point, or the remaining
subsequence ends with a run in the opposite direction of length at
least $k$. When $3\le s<k$, removal of $j$ reduces the final-run
length to $s-1$ without changing an earlier run. When $s=k$, the
final run before removal has exactly $k$ points or more than $k$
points, leaving $k-1$ or at least $k$ points. The length of the
remaining subsequence is therefore at most the corresponding value
in~\eqref{opt:weights}. Its endpoint $p$ satisfies the restrictions
in~\eqref{opt:recurrence}.

Conversely, consider a finite length in one of the maxima in
\eqref{opt:recurrence}. By the induction hypothesis and
\eqref{opt:weights}, there is a subsequence of that length ending at
$p$, or the singleton $p$ when that length is one. Append $j$.
The comparison with $x_p$ gives the required final-run direction,
and the bound on the segment indices preserves the segment
restriction. The cases of~\eqref{opt:weights} either start a
two-point run or extend the existing final run to the required
length. A change of direction uses $p$ as the shared turning point
and adds only the new point $j$.

These two implications prove the recurrence by induction on $j$.
The lengths $\dptab[j,k]{\sigma}$ are exactly the greatest lengths of
rollercoasters subject to the segment restriction. The locality lemma ensures
that this restriction leaves a longest rollercoaster available,
which proves the final assertion.
\end{proof}

\subsection{The data structure for monotone query subsequences}\label{opt:frontier-section}
For a fixed direction $\sigma$ and final-run length $s$, an earlier point $p$
contributes its value $x_p$ and the length $\dppred{\sigma}{p}{s}$
available before an extension. That is, a query at $x_j$ asks for the largest
of these lengths subject to $x_p<x_j$ when $\sigma=+$, or $x_p>x_j$
when $\sigma=-$. We prove a general data-structure lemma for the first
comparison; reversing value order gives the second. The lemma uses
the monotonicity of the query values along the subsequences of a segment.

A \emph{record} is a pair $(a,w)$, with key $a$ and
weight $w$. The initial $N_0$ records are supplied in increasing key
order. There are $N\ge1$ queries, each at a fresh key. A query at
$a$ returns the largest weight among records with key smaller than
$a$, together with a record of that weight, or $-\infty$ if there
is no such record. After the query, a record with the same key $a$
and a supplied weight may be inserted. A record's weight remains
fixed after insertion. Keys and weights can be compared in constant
time. In our application, the key is the point's value and the weight
is the length available before an extension, not the new
pseudocoaster length.

The queries are partitioned into $K$ nonempty subsequences of their
processing order. Within each query subsequence, the keys are strictly
increasing or strictly decreasing. The partition and the direction of
each subsequence are supplied; queries from different subsequences may
be interleaved arbitrarily.
All initial and query keys are distinct. In particular, $1\le K\le N$.
There is no assumed bound in terms of $K$ on the number of records.
The following lemma gives the required operation bounds.

\begin{lemma}\label{opt:frontier-lemma}
These operations take $O(N_0+N+N\log K)$ deterministic time and
$O(N_0+N+K)$ space. Every finite answer includes a record of the
returned weight; ties may be resolved arbitrarily.
\end{lemma}

\begin{proof}
We first explain which records are needed to answer queries. We
then describe the finger-search tree and the remembered search
position for each query subsequence. Finally, we bound both the searches and
the maintenance of those positions under deletions.

\subparagraph*{Retaining the records needed by later queries.}
A record $(a,w)$ can be discarded when another available record
$(b,v)$ has $b<a$ and $v\ge w$. Any query that could use the first
record can also use the second, with a weight at least as large.
The records left after all such deletions have weights strictly
increasing with their keys. Hence the answer to a query is the
weight of its \emph{predecessor}, the retained record with largest
key smaller than the query key. If there is no predecessor, the
answer is $-\infty$.

Suppose that a query at $a$ has returned predecessor $u$ and that a
record $(a,w)$ is to be inserted. If $u$ exists with weight at least
$w$, discard the new record. Otherwise insert it immediately after
$u$, or at the beginning when no predecessor exists. Delete the
consecutive records to its right with weights at most $w$, stopping
at the first larger weight or at the end. Because retained weights
are increasing, these are exactly the old records made redundant
by the insertion. This procedure maintains the required ordering
and leaves every query maximum unchanged.

Every supplied record is inserted at most once and deleted at most
once. Deleting a record from this search structure does not require
deleting other data that it referred to. In the algorithmic
application, links used to retrieve a subsequence will be retained
separately.

\subparagraph*{The tree and its search positions.}
Each remembered search position is stored as a pointer, called a
\emph{finger}, to a retained record from which a search starts.
We also keep two permanent artificial records with
keys $-\infty$ and $+\infty$, called \emph{sentinels}. They have no
weights and are never deleted or returned as query answers.
The left sentinel is used when a key has no
predecessor. The \emph{successor} of a query key is the record with
smallest larger key, or the right sentinel when no such record exists.

Store the records and sentinels in the level-linked weak B-tree of
Huddleston and Mehlhorn~\cite{HuddlestonM82}. We use their
$(2,4)$-tree with rebalancing Algorithm~2, which supports a search
from a given finger in $O(\log(d+2))$ time. Here $d$ is the number
of steps between the starting and ending positions in the current
ordered list. Insertion and deletion at known positions take
$O(1)$ amortised time. The distance is measured in the current
search structure.

A doubly linked list gives constant-time access to neighbouring
records. Each record is stored at a stable address and has a
pointer to its current tree leaf. A tree update that moves a record
between leaves updates this pointer, without moving the separately
stored record. Thus a tree rebalancing does not invalidate fingers,
and several query subsequences may have fingers at the same record.

The sorted initial records can be filtered in one scan: retain a
record exactly when its weight exceeds every preceding weight.
Build the tree by insertions at known positions in $O(N_0)$ total
time. Allocate one finger per query subsequence. Until the first query
in that subsequence, keep its finger at the left sentinel if its keys
increase, and at the right sentinel if they decrease.

Afterwards, maintain the following invariant: the finger lies strictly
below the last query key in an increasing subsequence, and strictly
above it in a decreasing subsequence. For the next query in an
increasing subsequence, search right from its finger for the predecessor
and move the finger there. For a decreasing subsequence, search left
for the successor and move the finger there; the successor's left
neighbour gives the predecessor needed by the query. Monotonicity of
the query keys and the invariant ensure that these searches have the
stated direction.

Perform this finger movement before the optional insertion at the
query key. The insertion does not move the finger onto the new
record. This distinction is necessary: the finger must remain
strictly on the stated side of the query subsequence's last query key.

\subparagraph*{Updating fingers after deletions.}
An insertion can delete a consecutive sequence of records that
contains the targets of several fingers. Redirect each increasing
finger on a deleted record to the newly inserted record, immediately
to the left of the deleted sequence. Redirect each decreasing finger
to the first record on its right that is not deleted, or to the
right sentinel. These replacements preserve the strict finger
invariant: the replacement key is smaller than the deleted key for
an increasing query subsequence and larger for a decreasing query subsequence.
Each redirection crosses only records deleted by this insertion;
it crosses no record that remains in the structure.

Figure~\ref{opt:fig:frontier-repair} illustrates both directions.

\begin{figure}[htbp]
\centering
\begin{tikzpicture}[x=1cm,y=1cm,every node/.style={font=\small},
  record/.style={draw,minimum width=1.30cm,minimum height=.55cm,
                 inner sep=3pt,fill=blue!5},
  repair/.style={->,thick}]
  \node at (5.5,.95) {Retained records in increasing key and weight order};
  \node[anchor=east] at (-.05,0) {Before};
  \node[anchor=east] at (-.05,-2.15) {After};

  \node[record] (b1) at (1.0,0) {$(1,2)$};
  \node[record] (b4) at (2.85,0) {$(4,3)$};
  \node[record] (b6) at (4.7,0) {$(6,6)$};
  \node[record,dashed,fill=red!10] (b9) at (8.9,0) {$(9,8)$};
  \node[record] (b12) at (11.0,0) {$(12,10)$};
  \draw[->] (b1.east) -- (b4.west);
  \draw[->] (b4.east) -- (b6.west);
  \draw[->] (b6.east) -- (b9.west);
  \draw[->] (b9.east) -- (b12.west);
  \node at (8.9,.51) {deleted};

  \node[record] (a1) at (1.0,-2.15) {$(1,2)$};
  \node[record] (a4) at (2.85,-2.15) {$(4,3)$};
  \node[record] (a6) at (4.7,-2.15) {$(6,6)$};
  \node[record,thick,fill=green!12] (a7) at (6.8,-2.15) {$(7,9)$};
  \node[record] (a12) at (11.0,-2.15) {$(12,10)$};
  \draw[->] (a1.east) -- (a4.west);
  \draw[->] (a4.east) -- (a6.west);
  \draw[->] (a6.east) -- (a7.west);
  \draw[->] (a7.east) -- (a12.west);
  \node at (6.8,-2.67) {inserted};

  \draw[repair] (b9.south west) -- (a7.north)
    node[midway,left=5pt,align=right]
      {$f_{\uparrow}$ repair\\to smaller key};
  \draw[repair] (b9.south east) -- (a12.north)
    node[midway,right=5pt,align=left]
      {$f_{\downarrow}$ repair\\to larger key};
\end{tikzpicture}
\caption{The retained records before and after inserting $(7,9)$, which deletes
$(9,8)$ because $7<9$ and $9\ge8$; the fingers $f_{\uparrow}$ and
$f_{\downarrow}$ on the deleted record are moved to $(7,9)$ and to
$(12,10)$, respectively.}
\label{opt:fig:frontier-repair}
\end{figure}
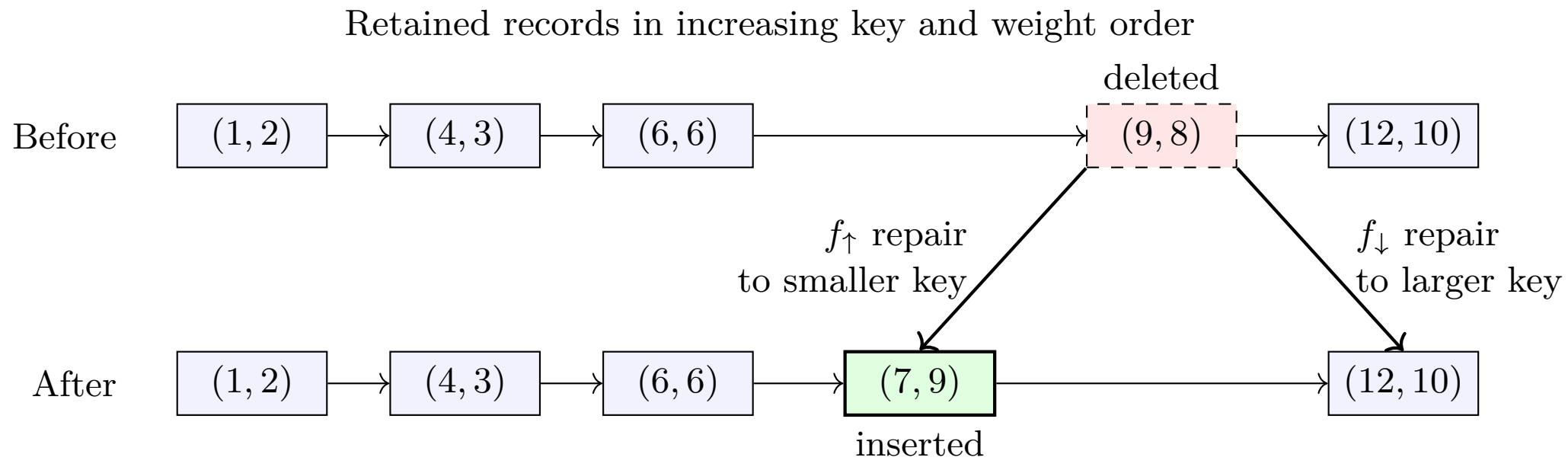
\FloatBarrier

\subparagraph*{Bounding the searches.}
For query $i$, let $d_i$ be the distance in the current ordered list
of retained records from its finger to the predecessor found by an
increasing search, or to the successor found by a decreasing search.
Count each step by the record entered; the starting record contributes
no step. The finger for an increasing query subsequence crosses any
one record to the right at most
once between that record's insertion and deletion, or termination.
Indeed, its queries move the finger right, and a deletion can move
it left only across records deleted in that operation. A record
already crossed therefore cannot be crossed again while it remains
in the structure. Insertions do not move fingers. The symmetric
argument applies to decreasing query subsequences.

There are at most $N_0+N$ inserted records and $K$ query subsequences.
Allowing constant work per query for endpoints and sentinels gives
\[
 \sum_{i=1}^{N}(d_i+2)=O\bigl(K(N_0+N)+N\bigr).
\]
The cost of search $i$ is $O(\log(d_i+2))$. Concavity of the binary
logarithm gives
\[
 \sum_{i=1}^{N}\log_2(d_i+2)
 \le N\log_2\!\left(\frac{1}{N}
                  \sum_{i=1}^{N}(d_i+2)\right).
\]
Substitution of the distance bound yields
\begin{align}
 \sum_{i=1}^{N}O(\log(d_i+2))
   &=O\!\left(N+N\log K+
                    N\log_2\left(1+\frac{N_0}{N}\right)\right)\notag\\
   &=O(N_0+N+N\log K).
   \label{opt:search-sum}
\end{align}
The first line uses $K(N_0+N)+N\le(K+1)(N_0+N)$ and
$\log_2(K+1)\le1+\log_2 K$ for $K\ge1$. The second uses
$\log_2(1+x)=O(x)$ for $x\ge0$. In particular, the bound includes
the cases $K=1$ and $N_0>N$.

Tree updates and traversing the deleted records take $O(N_0+N)$
total time. The remaining issue is the cost of redirecting fingers:
updating every finger separately after each deletion could exceed
the claimed bound.

\subparagraph*{Combining fingers with the same target.}
At each retained record, keep the increasing fingers in one set and
the decreasing fingers in a separate set. Each nonempty set has a
shared header containing its target,
size, and a doubly linked list of its members. A finger stores a
pointer to its header and its position in that member list.
Each search record points to its at most two headers.
Thus reading a finger's target takes constant time. Moving one
finger after a query also takes constant pointer work: detach it
from its old list and attach it to the destination list, creating
or removing a header if necessary.

For an insertion that deletes records, let the left replacement
be the newly inserted record and the right replacement the first
record after all deleted records, possibly the right sentinel.
Transfer the sets of increasing fingers from the deleted records to
the left replacement and the sets of decreasing fingers to the right.
If the destination has no fingers of that direction, only the
header's target and the record-to-header pointers change.
Otherwise combine the two sets: retain the larger set's header,
change every member of the smaller set to point to that header,
and splice the lists. Set the retained header's target to the
destination record, even when that header came from a deleted
record. Perform these transfers before freeing the deleted records.
Stable addresses for records and headers make these
operations independent of tree rebalancing.

There are at most two such transfers per deleted record. A transfer
to an empty destination takes constant time; combining two nonempty
sets of sizes $a\le b$ takes $O(a)$ time for changing member pointers.
We bound the total of these changes by the potential
\[
 \pot=\sum_G |G|\log_2|G|,
\]
where the sum ranges over the nonempty sets of fingers of both directions.
They partition the $K$ fingers, so $0\le\pot\le K\log_2 K$.
Combining sets of sizes $a\le b$ increases the potential by
\[
 (a+b)\log_2(a+b)-a\log_2a-b\log_2b\ge a.
\]
Indeed, the left side is
$a\log_2((a+b)/a)+b\log_2((a+b)/b)$, whose first term is at least
$a$ because $b\ge a$, and whose second term is nonnegative.
This increase pays for the $a$ changed member pointers. Detaching a
finger from a singleton set leaves the potential unchanged, since
that set contributes zero. For larger sets, the derivative of
$z\log_2 z$ bounds the decrease by $\log_2 K+O(1)$. Attaching that
finger to its new set cannot decrease the potential. Transferring
a whole set without combining it leaves the potential unchanged.

Let $R$ be the number of member-pointer changes during combinations,
and let $D$ be the total potential decrease due to detachments.
There is at most one detachment per query, hence
$D=O(N+N\log K)$. The only potential decreases are those counted
in $D$, and therefore
\begin{align*}
 R&\le\sum_{\text{combinations}}\Delta\pot
   \le\pot_{\mathrm{final}}-\pot_{\mathrm{initial}}+D\\
  &\le K\log_2 K+O(N+N\log K)
   =O(N+N\log K).
\end{align*}
The last equality uses $K\le N$. The other work on set
headers is $O(N_0+N)$. Combining this with~\eqref{opt:search-sum}
proves the time bound.

At most $N_0+N$ records are inserted. The tree and linked list use
linear space in the number of retained records, and the fingers
and their headers use $O(K)$ space. This proves the stated space
bound and completes the proof.
\end{proof}

\subsection{Computing the pseudocoaster lengths}\label{opt:evaluation}
We now compute the two length functions with the data structure of
Lemma~\ref{opt:frontier-lemma}. For each of the $2(k-1)$ pairs
$(\sigma,s)$, use a separate search structure. A point $p$ supplies a record
with key $x_p$ and weight $\dppred{\sigma}{p}{s}$ from
\eqref{opt:weights} when that weight is finite. Attach a link to
the pseudocoaster length, or the singleton, that equals the weight.
The data-structure weights are thus lengths of subsequences that
can be extended to compute the lengths for final-run length $s$.

For $\sigma=+$, use the ordinary value order; a query at $x_j$
then considers records with $x_p<x_j$. For $\sigma=-$, reverse the
comparison order in this structure. A smaller key in
that order means a larger input value, so the same query operation
considers records with $x_p>x_j$. This requires only comparisons,
not arithmetic on input values.

Each monotone subsequence of a segment supplies one query subsequence:
the queries generated by its points, in input order. Every point
generates one query in each of the $2(k-1)$ search structures, and
its queries inherit its membership in the segment's partition.
The list indices and directions were recorded during the decomposition.
There are at most $k$ nonempty query subsequences, interleaved in input
order. Reversing comparisons exchanges their increasing and decreasing
directions.

\subparagraph*{Initialising the structures for a segment.}
At the beginning of $P_\ell$, the lengths at all preceding positions
have been computed. Initialise each structure from the finite lengths
contributed by points in the preceding at most four segments,
\[
 P_{\max(1,\ell-4)}\cup\cdots\cup P_{\ell-1},
\]
with an empty initial set for $\ell=1$. Merge their sorted value
lists in time linear in their total size. This gives the records
in increasing key order for $\sigma=+$; reverse it for $\sigma=-$.
For each final-run length, omit records whose length is $-\infty$.
Allocate one finger for each nonempty query subsequence of $P_\ell$ as in
Appendix~\ref{opt:frontier-section}.

Reinitialising at segment boundaries is important. A record from
more than four preceding segments must no longer be available,
but the corresponding pseudocoaster lengths and backtracking links remain
stored. Inside a segment there is no removal because of age:
only records that cannot improve a query answer are discarded.

\subparagraph*{Processing a point.}
Process the positions $j$ of $P_\ell$ in input order. First query
all $2(k-1)$ structures at $x_j$, and add one to each answer to
obtain $\dptab[j,s]{\sigma}$. Next compute all lengths
$\dppred{\sigma}{j}{s}$ from the completed lengths at $j$, using
\eqref{opt:weights}. For each finite length, retain the link to
the pseudocoaster length or singleton that equals it and insert its record
in the corresponding structure. All queries at $j$ precede all
insertions at $j$, so no length at $j$ is computed from $j$ itself.

Each earlier point allowed by the segment restriction supplies its
value and finite length from~\eqref{opt:weights}; no other point
supplies a record. Discarding a record because a smaller key has at
least its stored length does not change any query maximum. Reversing the
key order for $\sigma=-$ gives exactly the opposite comparison
needed in~\eqref{opt:recurrence}. Each query therefore evaluates
the required recurrence, and Lemma~\ref{opt:dp-correct} proves
that the largest finite length $\dptab[j,k]{\sigma}$ is the optimum.

\subparagraph*{An example of a query and an insertion.}
Let $k=4$ and consider the structure computing the lengths
$\dptab[j,3]{+}$ for final-run length~3. The record at an earlier position $p$ stores the length
$\dppred{+}{p}{3}=\dptab[p,2]{+}$, since appending a larger value
extends the final increasing run from two points to three.
Suppose that the retained records are $(2,5)$, $(5,8)$, and
$(9,10)$. For a current value $x_j=7$, the comparison is
\[
\begin{array}{c|c|c}
 \text{key }x_p & \text{length }\dptab[p,2]{+} & x_p<7\\\hline
 2&5&\text{yes}\\
 5&8&\text{yes}\\
 9&10&\text{no}
\end{array}
\]
The record with key~5 has the largest permitted length, namely~8.
Append $7$ to the subsequence represented by that record, giving
$\dptab[j,3]{+}=9$, and retain a link to that pseudocoaster of length~8.
The new subsequence has nine points in total and three in its
final run. The record with key~9 cannot be used, despite its
larger stored length, because $9>7$.

After all queries at $j$, the record inserted into this same
structure stores the length $\dppred{+}{j}{3}=\dptab[j,2]{+}$, not the
value~9 just computed for final-run length~3. For example, if
$\dptab[j,2]{+}=8$, the new record is $(7,8)$. The existing
record $(5,8)$ has a smaller key and the same length, so the
new record is discarded. This does not delete the pseudocoaster length
$\dptab[j,3]{+}=9$ or its link for retrieving the subsequence.

\subsection{Backtracking and complexity}\label{opt:complexity}
\subparagraph*{Retaining the choices behind the pseudocoaster lengths.}
Store a permanent backtracking record for every singleton and
every finite pseudocoaster length. The record for $\dptab[j,s]{\sigma}$
contains the endpoint $j$ and a link to the earlier pseudocoaster length
or singleton that was extended by $j$. Once assigned, that link never
changes. The search records in the data structures refer to these
permanent records; deleting a search record leaves its target intact.

More explicitly, when the query for $\dptab[j,s]{\sigma}$ returns
a record of maximum stored length at an earlier position $p$, copy the
attached link into the backtracking record for the new length.
If $s=2$, that link identifies the singleton $p$ or the
opposite-direction length $\dptab[p,k]{\overline\sigma}$. For the other
final-run lengths, it identifies one of the lengths
in~\eqref{opt:weights}. After the lengths at $j$ have been
computed, each new search record links to the pseudocoaster length
or singleton that equals its stored weight. Ties may be resolved arbitrarily.

\subparagraph*{Retrieving the subsequence.}
Keep a link to a largest finite length $\dptab[j,k]{\sigma}$, over both
directions. Follow the backtracking links from this length, recording
each endpoint until reaching a singleton, then reverse the list.
Every link decreases the input position strictly. The list thus
contains at most $n$ positions and is obtained in $O(n)$ time.
The recurrence guarantees a rollercoaster of the recorded length.
At a direction change, the turning point is the endpoint of the
previous pseudocoaster length and is listed once. If all lengths
$\dptab[j,k]{\sigma}$ are $-\infty$, report nonexistence.

\subparagraph*{Total running time.}
Let $N_\ell=|P_\ell|$ and let $B_\ell$ be the total number of
points in its preceding at most four segments. For a fixed pair
$(\sigma,s)$, apply Lemma~\ref{opt:frontier-lemma} with
$N=N_\ell$, $N_0\le B_\ell$, and $1\le K\le k$. Since $k\ge3$,
the time for this structure on this segment is
\[
 O\bigl(B_\ell+N_\ell+N_\ell\log k\bigr)
   =O\bigl(B_\ell+N_\ell\log k\bigr).
\]
This includes initialisation, queries, insertions, and finger
maintenance. A point lies in one current segment and provides
initial records for at most four later segments, giving
\[
 \sum_\ell N_\ell=n,
 \qquad \sum_\ell B_\ell\le4n.
\]
The sum over all segments and all $2(k-1)$ pairs $(\sigma,s)$ of a
direction and a final-run length is therefore $O(nk\log k)$.

Evaluating~\eqref{opt:weights} and retaining the choices for the
pseudocoaster lengths take $O(nk)$ additional time. The decomposition,
its monotone lists, and sorting each segment take $O(n\log k)$.
Merging at most four sorted segment lists at each boundary takes
$O(n)$ in total. Finally, retrieving the subsequence takes $O(n)$.
All these costs fit within $O(nk\log k)$.

\subparagraph*{Space.}
The pseudocoaster lengths, their backtracking records, and the links associated
with the stored lengths use $O(nk)$ space. On a segment, one search
structure uses $O(B_\ell+N_\ell+k)$ space. The $2(k-1)$ structures therefore
use $O(nk+k^2)=O(nk)$ space, since $k\le n$. Discard those search
structures at the next segment boundary while retaining the
pseudocoaster lengths and backtracking links. The decomposition, sorted lists, and
output require $O(n)$ further space. This completes the proof of
Theorem~\ref{opt:thm:longest}.

A longest rollercoaster can therefore be found in
$O(n\min\{k\log k,\log^2 n\})$ time by choosing between our algorithm
and the $O(n\log^2 n)$-time algorithm of~\cite{GawrychowskiMS22}.

If only the guaranteed length is required, write $\extremalf{k}{n}$
for the minimum longest-rollercoaster length over all $n$-point
permutations. The algorithm of Appendix~\ref{app:algorithm} finds a
rollercoaster of at least that length in
$O(n\log\extremalf{k}{n})$ time for $n>(k-1)^2$.
Choosing between that algorithm and the longest-rollercoaster algorithm
proved here gives the following bound.
\begin{corollary}\label{opt:guarantee}
For $n>(k-1)^2$, a $k$-rollercoaster of length at least $\extremalf{k}{n}$ can be found
deterministically in $O(n\min\{k\log k,\log \extremalf{k}{n}\})$ time.
\end{corollary}
The guarantee and exact algorithms use $O(n)$ and $O(nk)$ space, respectively.

%% file: rollercoasters.bib
@book{Kitaev11,
  author = {Sergey Kitaev},
  title = {Patterns in Permutations and Words},
  series = {Monographs in Theoretical Computer Science. An EATCS Series},
  publisher = {Springer},
  year = {2011}
}

@article{Dilworth50,
  author = {Robert Palmer Dilworth},
  title = {A Decomposition Theorem for Partially Ordered Sets},
  journal = {Annals of Mathematics},
  volume = {51},
  number = {1},
  pages = {161--166},
  year = {1950}
}

@article{BiedlBCLMNS19,
  author = {Therese Biedl and Ahmad Biniaz and Robert Cummings and Anna Lubiw and Florin Manea and Dirk Nowotka and Jeffrey Shallit},
  title = {Rollercoasters: Long Sequences without Short Runs},
  journal = {SIAM Journal on Discrete Mathematics},
  volume = {33},
  number = {2},
  pages = {845--861},
  year = {2019}
}

@article{GawrychowskiMS22,
  author = {Pawe{\l} Gawrychowski and Florin Manea and Rados{\l}aw Serafin},
  title = {Fast and Longest Rollercoasters},
  journal = {Algorithmica},
  volume = {84},
  number = {4},
  pages = {1081--1106},
  year = {2022}
}

@inproceedings{BiedlCDJL18,
  author = {Therese Biedl and Timothy M. Chan and Martin Derka and Kshitij Jain and Anna Lubiw},
  title = {Improved Bounds for Drawing Trees on Fixed Points with {L}-Shaped Edges},
  booktitle = {Proceedings of the 25th International Symposium on
  Graph Drawing and Network Visualization (GD)},
  pages = {305--317},
  year = {2018},
  note = {Full version: arXiv:1709.01456 (2017)}
}

@inproceedings{BiedlBCLMNS18,
  author = {Therese Biedl and Ahmad Biniaz and Robert Cummings and Anna Lubiw and Florin Manea and Dirk Nowotka and Jeffrey Shallit},
  title = {Rollercoasters and Caterpillars},
  booktitle = {Proceedings of the 45th International Colloquium on Automata, Languages, and Programming (ICALP)},
  pages = {18:1--18:15},
  year = {2018}
}

@inproceedings{GawrychowskiMS19,
  author = {Pawe{\l} Gawrychowski and Florin Manea and Rados{\l}aw Serafin},
  title = {Fast and Longest Rollercoasters},
  booktitle = {Proceedings of the 36th International Symposium on Theoretical Aspects of Computer Science (STACS)},
  pages = {30:1--30:17},
  year = {2019}
}

@article{ErdosS35,
  author = {Paul Erd{\H o}s and George Szekeres},
  title = {A combinatorial problem in geometry},
  journal = {Compositio Mathematica},
  volume = {2},
  pages = {463--470},
  year = {1935}
}

@article{Seidenberg59,
  author = {Abraham Seidenberg},
  title = {A simple proof of a theorem of {Erd{\H o}s} and {Szekeres}},
  journal = {Journal of the London Mathematical Society},
  volume = {s1-34},
  number = {3},
  pages = {352},
  year = {1959}
}

@article{Chung80,
  author = {Fan Rong K. Chung},
  title = {On unimodal subsequences},
  journal = {Journal of Combinatorial Theory, Series A},
  volume = {29},
  number = {3},
  pages = {267--279},
  year = {1980}
}

@article{AhmedS13,
  author = {Tanbir Ahmed and Hunter Snevily},
  title = {Some properties of roller coaster permutations},
  journal = {Bulletin of the Institute of Combinatorics and its Applications},
  volume = {68},
  pages = {55--69},
  year = {2013}
}

@inproceedings{BulteauFJV22,
  author = {Laurent Bulteau and Guillaume Fertin and Vincent Jug{\'e} and St{\'e}phane Vialette},
  title = {Permutation Pattern Matching for Doubly Partially Ordered Patterns},
  booktitle = {Proceedings of the 33rd Annual Symposium on Combinatorial Pattern Matching (CPM)},
  pages = {21:1--21:17},
  year = {2022}
}

@article{DiStefanoKLZ08,
  author = {Gabriele {Di Stefano} and Stefan Krause and Marco E. L{\"u}bbecke and Uwe T. Zimmermann},
  title = {On minimum {$k$}-modal partitions of permutations},
  journal = {Journal of Discrete Algorithms},
  volume = {6},
  number = {3},
  pages = {381--392},
  year = {2008}
}

@article{HuddlestonM82,
 author = {Scott Huddleston and Kurt Mehlhorn},
 title = {A New Data Structure for Representing Sorted Lists},
 journal = {Acta Informatica},
 volume = {17},
 number = {2},
 pages = {157--184},
 year = {1982}
}

@inproceedings{AdamsonFH24,
 author = {Duncan Adamson and Pamela Fleischmann and Annika Huch},
 title = {Rollercoasters with Plateaus},
 booktitle = {Proceedings of the 18th International Conference
 on Reachability Problems (RP)},
 pages = {73--87},
 year = {2024}
}

@inproceedings{FujitaNIBT21,
 author = {Kosuke Fujita and Yuto Nakashima and Shunsuke Inenaga and Hideo Bannai and Masayuki Takeda},
 title = {Longest Common Rollercoasters},
 booktitle = {Proceedings of the 28th International Symposium
 on String Processing and Information Retrieval (SPIRE)},
 pages = {21--32},
 year = {2021}
}

@article{AldousD99,
  author = {David Aldous and Persi Diaconis},
  title = {Longest increasing subsequences: from patience sorting to the {Baik--Deift--Johansson} theorem},
  journal = {Bulletin of the American Mathematical Society},
  volume = {36},
  number = {4},
  pages = {413--432},
  year = {1999}
}
